\documentclass[sigplan,10pt]{acmart}

\renewcommand\footnotetextcopyrightpermission[1]{} % removes footnote with conference information in first column
\usepackage{booktabs}   %% For formal tables:
\usepackage{caption}
\usepackage[font=footnotesize]{subfig}
\usepackage[ruled,vlined,linesnumbered]{algorithm2e}
\usepackage{enumitem}
\SetKwProg{Fn}{Function}{}{end}\SetKwFunction{construct}{RouteConstruct}%

\begin{document}
	
% \copyrightyear{2018}
% \acmYear{2018}
% \setcopyright{acmlicensed}
% \acmConference[PPoPP '18]{PPoPP '18: 23nd ACM SIGPLAN Symposium on Principles and Practice of Parallel Programming}{February 24--28, 2018}{Vienna, Austria}
% % \acmBooktitle{PPoPP '18: PPoPP '18: 23nd ACM SIGPLAN Symposium on Principles and Practice of Parallel Programming, February 24--28, 2018, Vienna, Austria}
% \acmPrice{15.00}
% \acmDOI{10.1145/3178487.3178491}
% \acmISBN{978-1-4503-4982-6/18/02}

% \copyrightyear{2018}
% \acmYear{2018}
% \setcopyright{acmlicensed}
% \acmConference[PPoPP '18]{PPoPP '18: 23nd ACM SIGPLAN Symposium on Principles and Practice of Parallel Programming}{February 24--28, 2018}{Vienna, Austria}
% \acmBooktitle{PPoPP '18: PPoPP '18: 23nd ACM SIGPLAN Symposium on Principles and Practice of Parallel Programming, February 24--28, 2018, Vienna, Austria}
% \acmPrice{15.00}
% \acmDOI{10.1145/3178487.3178491}
% \acmISBN{978-1-4503-4982-6/18/02}

%% Title information

\title[Short Title]{SuperNeurons: Dynamic GPU Memory Management for Training Deep Neural Networks}

%% Author information
%% Contents and number of authors suppressed with 'anonymous'.
%% Each author should be introduced by \author, followed by
%% \authornote (optional), \orcid (optional), \affiliation, and
%% \email.
%% An author may have multiple affiliations and/or emails; repeat the
%% appropriate command.
%% Many elements are not rendered, but should be provided for metadata
%% extraction tools.

%% Author with single affiliation.
\author{Linnan Wang \textsuperscript{{\normalsize$\S$}}
		Jinmian Ye  \textsuperscript{{\normalsize$\dagger$}} 
		Yiyang Zhao \textsuperscript{{\normalsize$\dagger$}} 
		Wei Wu      \textsuperscript{{\normalsize$\ddagger$}} 
		Ang Li      \textsuperscript{{\normalsize$\amalg$}}
		Shuaiwen Leon Song \textsuperscript{{\normalsize$\amalg$}} \\
		Zenglin Xu  \textsuperscript{{\normalsize$\dagger$}} 
		Tim Kraska  \textsuperscript{{\normalsize$\diamondsuit$}}  \\
        \textsuperscript{{\normalsize$\S$}} Brown University \\ \vspace{-0.18in}      		\textsuperscript{{\normalsize$\dagger$}} Univ. of Electr. Sci. \& Tech. of China \\
		\textsuperscript{{\normalsize$\ddagger$}} Los Alamos National Laboratory \\ \vspace{-0.18in}
		\textsuperscript{{\normalsize$\amalg$}} Pacific Northwest National Laboratory \\ 
        \textsuperscript{{\normalsize$\diamondsuit$}}  Massachusetts Institute of Technology\\
}

\begin{abstract}

Going deeper and wider in neural architectures improves the accuracy, while the limited 
GPU DRAM places an undesired restriction on the network design domain.
Deep Learning (DL) practitioners either need change to less desired network architectures, 
or nontrivially dissect a network across multiGPUs. These distract DL practitioners from concentrating
on their original machine learning tasks.
We present SuperNeurons: a dynamic GPU memory scheduling runtime to enable the network training 
far beyond the GPU DRAM capacity. SuperNeurons features 3 memory optimizations, 
\textit{Liveness Analysis}, \textit{Unified Tensor Pool}, and \textit{Cost-Aware Recomputation}; 
all together they effectively reduce 
the network-wide peak memory usage down to the maximal memory usage among layers. 
We also address the performance issues in those memory saving techniques. Given the limited GPU DRAM, 
SuperNeurons not only provisions the necessary memory for the training, but also dynamically allocates 
the memory for convolution workspaces to achieve the high performance.
Evaluations against Caffe, Torch, MXNet and TensorFlow have demonstrated that SuperNeurons trains 
at least 3.2432 deeper network than current ones with the leading performance. Particularly, 
SuperNeurons can train ResNet2500 that has $10^4$ basic network layers on a 12GB K40c.

\end{abstract}

  %% \keywords is optional

\maketitle

\section{Introduction}

Deep Neural Network (DNN) is efficient at modeling complex nonlinearities thanks to the unparalleled
representation power from millions of parameters.  
This implies scaling up neural networks is an effective approach to 
improve the generalization performance. The Deep Learning (DL) community now widely acknowledges 
either going deeper or going wider on the nonlinear architecture improves the quality
of image recognition tasks. For example, 9-layer AlexNet won the 2012 ILSVRC (ImageNet Large-Scale Visual Recognition 
Challenge) with a top-5 error of 17\%. GoogLeNet (inception v1) refreshed the top-5 error rate to 6.67\% 
with 22 inception units in 2014 ILSVRC, and ResNet further reduced the error rate down to 3.57\% 
in 2015 ILSVRC with 152 residual units.

While DL practitioners are enthusiastically seeking deeper and wider nonlinear networks, 
the limited size of GPU DRAM becomes a major restriction. Training a deep network is inherently a 
computing-intensive task. Almost every AI lab today, either in academia or industry, is deploying
the network training on GPUs for the demand of high-performance \cite{bahrampour2016comparative}. 
Data need to be residing on GPU DRAM for the GPU computing, but the largest commercial GPU DRAM so far 
is 24 GB. This is still far from sufficient to accommodate a deep neural network. For example, 
the latest Inception v4 has 515 basic layers consuming 44.3 GB memory in the training.
The deeper or wider we go, the higher memory usages will be.
Therefore, this deep trend subjects the rigid GPU DRAM to the severe insufficiency.

Major DL frameworks, such as Caffe or MXNet, have tried to alleviate the GPU memory shortage 
with several static memory reduction techniques. Those techniques, due to their static nature, 
are not well tuned to address the new data and dependency variations in non-linear 
networks. For example, Caffe and Torch do not fully support the data flow analysis on
non-linear neural networks; the trading computation for memory strategy
in MXNet is limited for ignoring the memory variations across network layers. 
These limitations have motivated us to propose a dynamic approach for the 
emerging deep nonlinear neural architectures.

In this paper, we present the first dynamic GPU memory scheduling runtime for 
training deep non-linear neural networks. The runtime allows DL practitioners to explore a
much deeper and wider model beyond the physical limitations of GPU memory. It utilizes
tensors as the fundamental scheduling units to consist with the layer-wise computations enforced in
DL performance primitives cuDNN \cite{chetlur2014cudnn}.
The runtime seamlessly orchestrates the tensor placement, movement, allocation and deallocation
so that the underly memory operations are entirely transparent to users.

Our runtime guarantees the minimal peak memory usage, $peak_m = \max(l_i)$, at the layer-wise granularity.
We denote the memory usage of the $ith$ layer as $l_i$, and the superscript, e.g. $l_i^{f}$ or $l_i^{b}$, as the forward/backward. The peak memory usage during the forward and backward computations is denoted as $peak_m$. 
First, \textit{Liveness Analysis} recycles no longer needed tensors 
to reduce $peak_m$ from baseline $\sum_{i=1}^{N} l_i^f + \sum_{i=1}^{N} l_i^b$ 
to $\sum_{i=1}^{N} l_i^f + l_{N}^{b}$ (defined in Sec.\ref{method}).
Secondly, Unified Tensor Pool (UTP) offloads tensors in compute-intensive layers, referred to as checkpoints, to the external 
physical memory. This further reduces $peak_m$ from $\sum_{i=1}^{N} l_i^f + l_{N}^{b}$ to 
$\sum_{i=1}^{N} (l_i^f \notin checkpoints) + l_{N}^{b}$. Finally, \textit{Cost-Aware Recomputation} drops the forward
results of cheap-to-compute or none-checkpoints layers and reconstructs them to reduce $peak_m$ from $\sum_{i=1}^{N} (l_i^f \notin checkpoints) + l_{N}^{b}$ to $peak_m = \max(l_i)$. The final $peak_m$ indicates the largest computable network is bounded by the maximal memory usage among layers.

Our runtime also features three performance optimizations to improve the efficiency of \textit{Liveness Analysis} and \textit{UTP}.
First, GPUs require memory allocations to create tensors and deallocations to free tensors. Thus, the high-frequent large tensor allocations/deallocations incur the non-negligible overhead in \textit{Liveness Analysis} \cite{wang2016blasx}. The runtime successfully amortizes the cost by directly reusing memory segments from a huge pre-allocated memory pool, managed by a heap based GPU memory management utility. Secondly, UTP swaps tensors among different physical memory spaces, while modern GPUs equip with independent Direct Memory Access (DMA) engine exposing opportunities to hide communications under computations. The runtime also meticulously overlap communications with computations. However, the overlapping opportunity is limited given the fixed amount of computations. We propose a LRU based Tensor Cache built on GPU DRAM to minimize total communications by tensor reusing.

This paper claims the following contributions:
\begin{itemize}[leftmargin=*]
  \item We demonstrate the new memory scheduling challenges in nonlinear neural networks, and discuss the key limitations of existing approaches.
  \item We design and implement SuperNeurons to enable DL practitioners to explore deep neural networks; and the largest computable network of SuperNeurons is only bounded by the max memory usage among layers.
  \item By dynamically allocating memory for convolution workspaces, SuperNeurons deliveries the leading performance among state-of-art DL systems on the GPU.
\end{itemize}

\section{Background and Motivation}
%This section starts with reviewing the GPU memory management in mainstream ML frameworks 
%and their key limitations. Then we present an exposition of memory scheduling challenges 
%in non-linear Convolutional Neural Networks (CNN), and elaborate several observations 
%that motivate the key techniques designed in our runtime.
%
\subsection{Challenges for Processing Super Deep Neural Networks}

\begin{figure}[t]
\subfloat[][fan]{\includegraphics[height=1.0in]{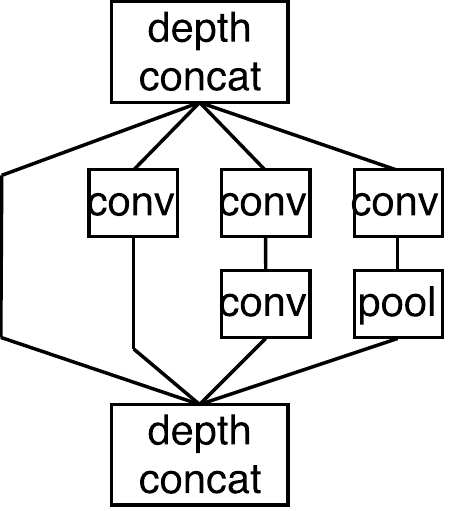}\label{fanConn}} \quad
\subfloat[][join]{\includegraphics[height=1.0in]{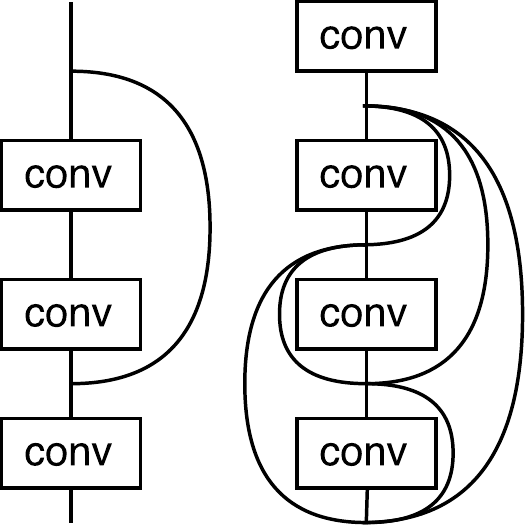}\label{joinConn}}  \quad
\caption{The non-linear connections in inception v4 (fan), ResNet (join, left) and DenseNet 
(join, right). DenseNet utilizes a full-join.}
\label{nonlinear_connections}
\vspace{-0.5cm}
\end{figure}

Traditional Convolutional Neural Networks (CNN)
\cite{ lecun1998gradient, krizhevsky2012imagenet, simonyan2014very} are typically composed 
of several basic building layers, including Convolution (CONV), Pooling (POOL), Activation (ACT), Softmax, Fully Connected (FC), 
Local Response Normalization (LRN), Batch Normalization (BN), and Dropout. For linear CNNs, 
these layers are independent and inter-connected to their neighbors in a sequential manner: 
$1 \leftrightarrow 2 \leftrightarrow \cdot\cdot\cdot \leftrightarrow n$. 
Recently, several deep non-linear neural architectures have been proposed to further 
improve the state-of-the-art accuracy on the 1K ImageNet recognition challenge, e.g., 
Inception v4\cite{szegedy2017inception}, ResNet\cite{he2016deep}, and DenseNet\cite{huang2016densely}. 
These prominent network designs (especially the one that solves classic gradient vanishing \cite{bengio1994learning} 
problem) pave the algorithmic foundation for DL practitioners to harness the unparalleled 
representation power brought forth by the super deep non-linear neural architectures. 
For example, the latest inception v4 delivers 95\% top-5 accuracy with 515 basic building 
layers while ResNet151\footnote{151 represents the number of convolutional units.} achieves 
94.3\% top-5 accuracy with 567 layers. In Figure \ref{nonlinear_connections}, we illustrate 
two classic types of non-linear connections: fan and join. Compared with the linear connection 
pattern, the sparse fan-out connection (Figure \ref{fanConn}) avoids one huge computing-inefficient 
dense layer \cite{szegedy2015going} while the join connection prevents gradients from quickly 
vanishing in the back-propagation \cite{he2016deep}. 

Training these super deep and complex non-linear neural architectures is a 
computation-intensive task. Due to its DL-driven novel architecture designs and massive 
parallelism, GPUs have been widely adopted in today's industry and academia for the efficient 
neural network training. However, there are critical issues for 
efficiently training in these newly developed super deep non-linear neural architectures:
\textit{limited GPU resident memory} and \textit{a high degree of variation in computational dependencies}. 

\begin{figure}[!t]
\centering
\includegraphics[width=0.75\columnwidth]{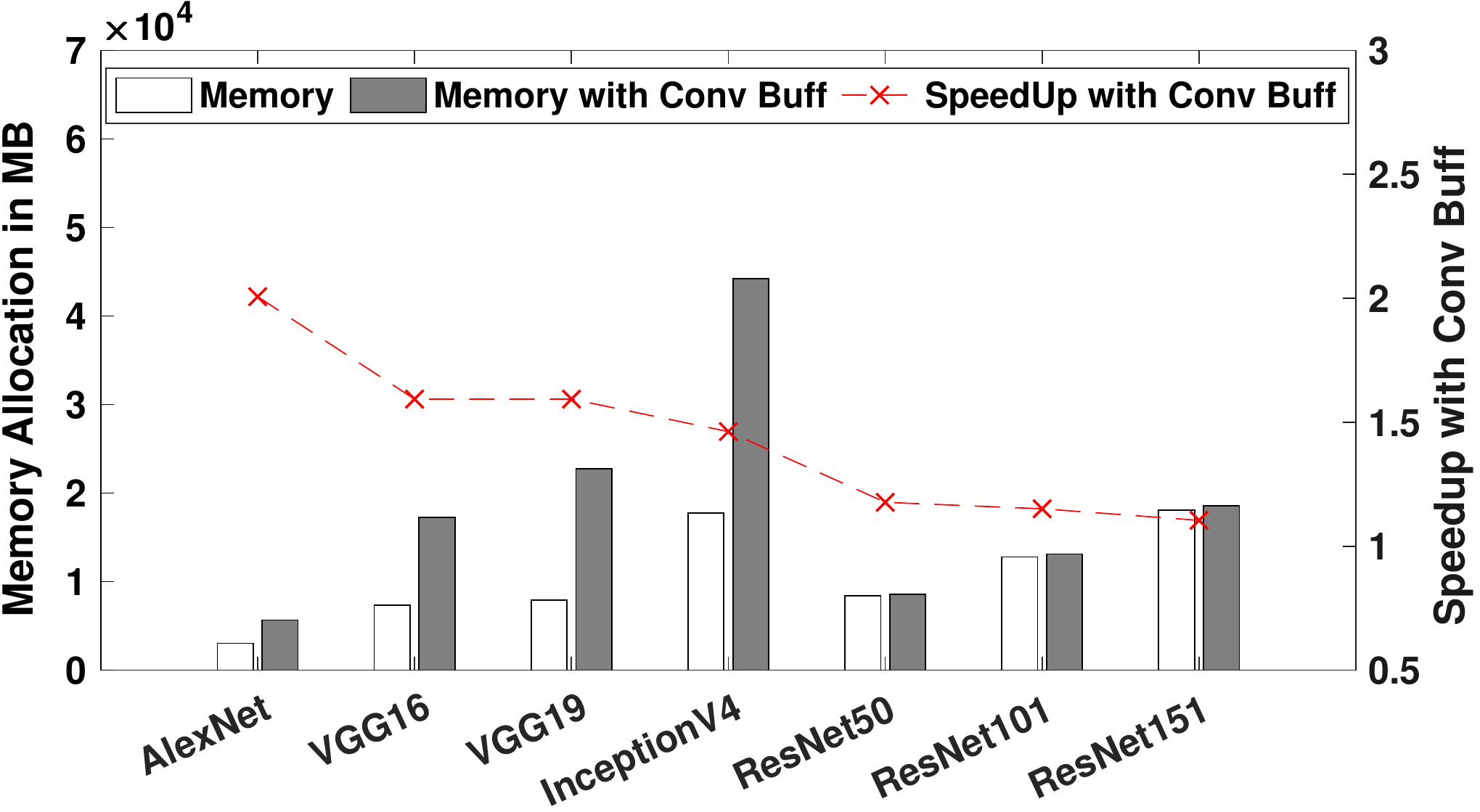}
\caption{The left axis depicts the memory usages of networks. The batch size of AlexNet 
is 200, and the rest use 32. The right axis and red x marks depict the speedup (imgs/s) with and without convolution workspaces. }
\label{figure1}
\vspace{-0.2cm}
\end{figure}

\textbf{Challenge I: Limited GPU Resident Memory.} The prominent deep neural architectures 
share a common feature: high memory demand and computation intensity. Figure \ref{figure1} 
illustrates the network wide memory usages of several recent DNNs in the training 
with and without convolution workspaces or buffer. Among them, AlexNet and VGG are linear networks while 
the others are non-linear.  We can observe that the non-linear networks demand a significant 
amount of GPU memory, e.g., ResNet152 and Inception v4 require up to 18.5GB and 44.3 GB at only the
batch size of 32, respectively. However, these 
sizes are either similar to or surpass the resident memory sizes of commercial GPUs on the 
market today. For instance, the newest generations of NVIDIA Pascal and Volta GPUs only have 
16GB with HBM2 enabled (e.g., P100 and V100) while the one with the most memory available in 
the recent generations is Maxwell P40 with 24GB GDDR5.  This limitation poses a major 
bottleneck for deep learning practitioners for exploring deep and wide neural architectures 
\cite{szegedy2015going, pleiss2017memory, szegedy2017inception}. The most straight forward 
solution is to split the network across GPUs, i.e. Model Parallelism. However, splitting 
either the computations of a network or a layer incurs excessive intra-network and intra-layer 
communications that drastically deteriorate the performance. For example, recent work has 
suggested the deficiency of applying model parallelism for deep neural networks: it compromises 
at least 40$\%$ speed when training a network with 1.3 billion parameters from 36 GPUs to 64 GPUs 
\cite{coates2013deep}. To address the performance issues from Model Parallelism, Data Parallelism 
has been widely adopted in today's mainstream deep learning frameworks such as 
Caffe\cite{jia2014caffe}, TensorFlow\cite{abadi2016tensorflow}, Torch\cite{collobert2002torch}, 
and MXNet\cite{chen2015MXNet}. In this model, each GPU holds a network replica; and one GPU 
computes one sub-gradient with a sub-batch. Subsequently all sub-gradients are aggregated 
as one global gradient to update the network parameters \cite{wang2016efficient}. Although 
this process does not incur intra-network or intra-layer communications besides necessary
gradient exchanges, it requires the network training to fit in the limited GPU 
DRAM. In this paper, we focus on addressing the GPU memory 
shortage issue for training deep neural networks under data parallelism model while taking the
training performance into design considerations.

\begin{figure*}[!t]
\centering
\subfloat[][linear]{\includegraphics[width=0.6\columnwidth]{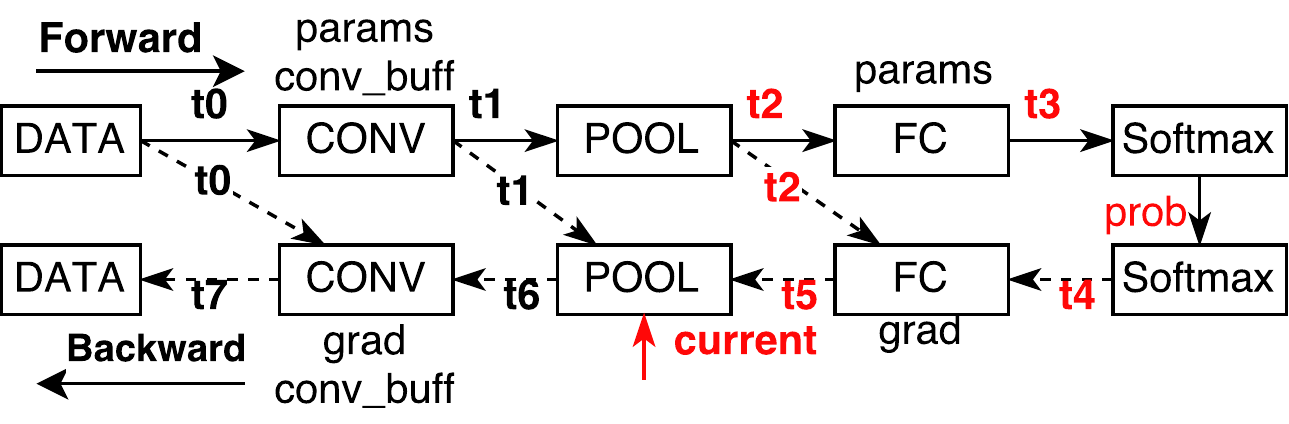}\label{linear_dep}} \quad
\subfloat[][join (nonlinear)]{\includegraphics[width=0.6\columnwidth]{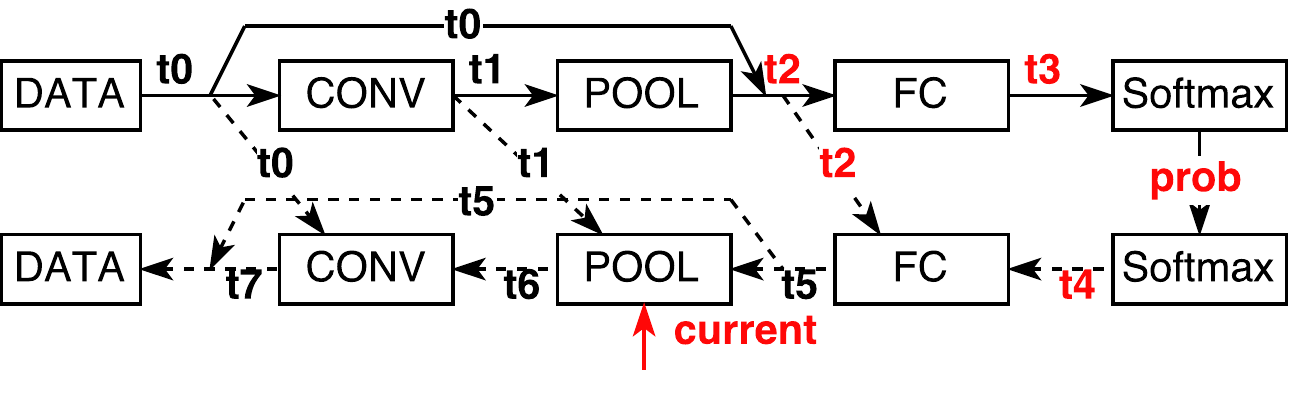}\label{join_dep}}  \quad
\subfloat[][fan (nonlinear)]{\includegraphics[width=0.6\columnwidth]{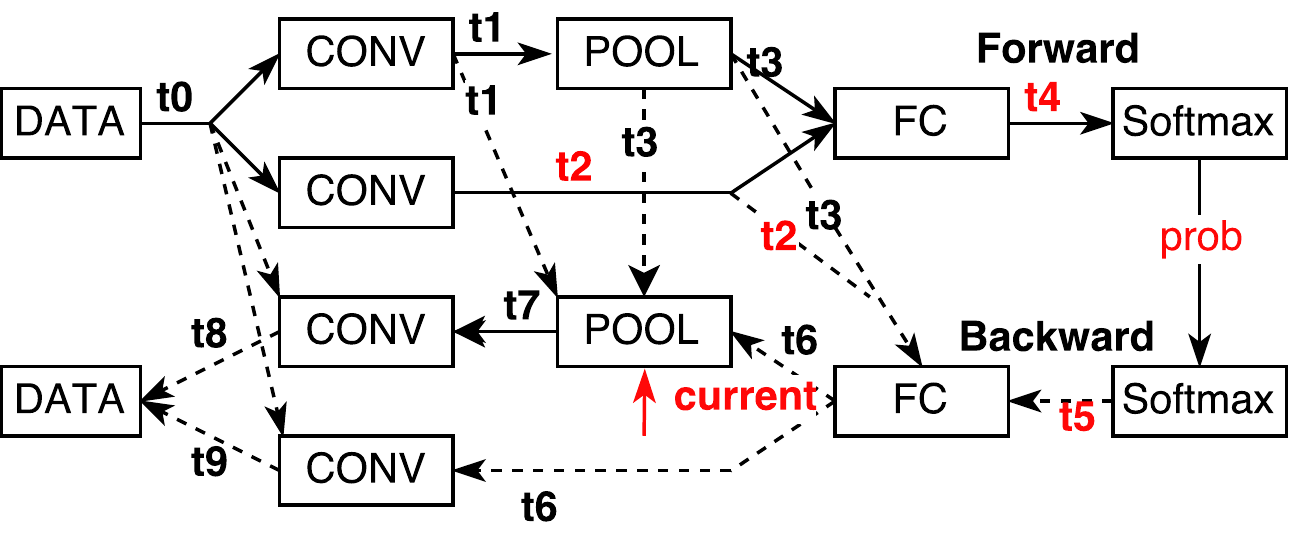}\label{fan_dep}}  \quad
\caption{Data dependencies of different neural architectures. Tensors in red are ready to free when back propagate to the POOL layer. 
Solid lines represent forward dependencies and dash lines represent backward dependencies.}
\label{fundamental_connections}
\vspace{-0.2cm}
\end{figure*}

\textbf{Challenge II: Variations in Computational Dependencies for Nonlinear Networks.} 
Nonlinear networks exhibit a high degree of dependency variations while linear networks follow 
a fixed sequential execution pattern with predictable data dependencies \cite{rhu2016vdnn}. 
Fig.\ref{fundamental_connections} illustrates the data dependency graph for linear (a) and nonlinear 
(b and c) neural architectures. One typical training iteration consists of two phases: 
forward and backward propagation. For Linear networks, data is sequentially propagated in 
the forward pass; and a layer's backward computation is simply contingent upon the previous 
layer as illustrated in Figure \ref{linear_dep}. Thus their computation and dependency 
patterns are static regardless of the total layers involved. 

However, for nonlinear networks, 
a high degree of variations in computational dependencies appear. Fig.\ref{join_dep} and 
\ref{fan_dep} show two simple examples of join and fan nonlinear connections. For join 
connections, it forwards a layer's output tensor to another layer, 
creating a dependency between two layers. For example, the join in Fig.\ref{join_dep} forwards 
$\mathbf{t0}$ from DATA layer to FC layer in the forward pass. The dependency of 
join-based non-linear networks is non-deterministic as any two layers can be connected with 
a join, e.g., in DenseNet. For fan connections, it creates multiple branches in the execution 
flow: DATA layer forks two branches and joins them before FC layer. Separate branches, 
each with a different number of layers, have to finish before joining them back to the original 
branch, making this execution sequence nonlinear. Although the two basic nonlinear 
scenarios shown here are intuitive, a typical deep nonlinear network today has hundreds of 
joins and fans convoluted together, resulting in a complex network architecture. These 
significantly complicate runtime resource-management compared to the static computational pattern in linear ones.
Therefore, the memory scheduling of deep non-linear neural networks demands a dynamic solution to effectively address 
these variations in both the execution flow and computation dependencies.

\subsection{Limitations of GPU Memory Management in Mainstream Deep Learning Frameworks}
%Mainstream DL frameworks, such as Caffe\cite{jia2014caffe}, 
%TensorFlow\cite{abadi2016tensorflow}, Torch\cite{collobert2002torch}, and MXNet\cite{chen2015MXNet}, 
%parallelize the training with Data Parallelism for performance that requires each GPU to hold a network replica,
%while the rigid 16 GB GPU DRAM not only fails to satisfy the network training, 
%but also subjects the training to low performance as suggested in Fig.\ref{figure1}.
%Several static memory management techniques have been discussed and implemented in 
%those frameworks to address the GPU memory shortage.

Several static memory reduction techniques have been implemented in today's deep learning 
frameworks to address the GPU memory shortage at data parallelism level. For example, Caffe 
and Torch directly reuse the forward data tensors for the backward data propagation, which 
saves up to $50\%$ of memory on a linear network \cite{MXNet_graph}. Although this 
technique works well on linear networks, it requires extra 
tensors to hold the future dependencies for training non-linear networks, thereby limiting 
the effectiveness and efficiency. Also, these frameworks still have to fit the entire 
network into GPU DRAM without leveraging NUMA architectures; 
and this level of reuse is arguably not adequate for contemporary deep 
nonlinear neural networks. MXNet and TensorFlow are built with a Directed Acyclic Graph (DAG) 
execution engine \cite{wu2015hierarchical}. Users explicitly define the computation flow 
and tensor dependencies, which provide necessary information for the DAG engine to analyze 
the life span of tensors. Both systems then free tensors that are no longer needed in order 
to save memory. MXNet also implements a per-layer-based re-computation strategy that is similar 
to Resilient Distributed Datasets (RDD) in Spark. Basically it frees 
the tensors produced by computing-cheap layers
in the forward pass, and recomputes the freed dependencies for the backward pass by doing 
another forward. However, this method neglects non-uniform 
memory distribution of network layers, consequentially demanding large unnecessary memory 
usages. TensorFlow swaps long-lived data tensors from GPU DRAM to CPU DRAM, but it fails 
to optimize data communications between the two (e.g., utilizing pinned data transfer) which 
compromises at least $50\%$ of communication speed.

%Several static memory reduction techniques have been implemented in DL frameworks to 
%address the GPU memory shortage at the Data Parallelism scheme. 
%
%
%Caffe and Torch 
%reuse tensors in the forward and backward data propagation\cite{MXNet_graph}; and this
%technique saves up to $50\%$ of memory on a linear network. MXNet and TensorFlow are built 
%with a Direct Acyclic Graph (DAG) execution engine \cite{wu2015hierarchical}. Users explicitly 
%define the computation flow and tensor dependencies, which provide necessary information 
%for the DAG engine to analyze the life span of tensors. Thereby both systems free no longer
%needed tensors to save memory. MXNet also utilizes a re-computation technique to reconstruct
%backward dependencies with another forward pass, while TensorFlow offloads long lived tensors 
%to CPU DRAM.

More importantly, none of aforementioned DL frameworks utilize a dynamic scheduling policy 
that provisions necessary memory space for deep nonlinear network training while at the 
same time optimizing the training speed given the existing GPU DRAM resource. In other words, 
these static memory-saving techniques aggressively reduce the GPU memory usage at the 
expense of speed. Users either painstakingly tune the performance or suffer from the 
insufficient memory during the execution. Additionally, these frameworks either have no optimization 
strategy or adopt a naive method on allocating the convolution workspace (see Section \ref{workspace}), 
which is a decisive factor determining CNN training speed on the GPU. In summary, these challenges 
motivate us to design a dynamic scheduling runtime to provision necessary memory for the training
while maximizing the memory for convolution workspaces to optimize the training speed. 

\section{Design Methodologies}

\label{method}
This section elaborates on three memory optimization techniques and their related performance issues in 
SuperNeurons. From a high level perspective, SuperNeurons provisions necessary memory spaces 
for the training while maximizing the speed by seeking convolution workspaces within the constraint 
of native GPU memory size.

\textbf{Notations and Baseline Definition:} To facilitate the analysis of proposed techniques, we denote the forward memory usage of the $ith$ layer as $l_i^{f}$, the backward as $l_i^{b}$. We denote the peak memory usage as $peak_m$. We use the naive network-wide tensor allocation strategy as the baseline, which allocates an independent tensor for each memory requests. Thus, the $peak_m$ of baseline is $\sum_{i=1}^{N} l_i^f + \sum_{i=1}^{N} l_i^b$. We also denote the maximal memory usage among layers as $l_{peak} = max(l_i)$, where $i \in [1, N]$, and $N$ represents the network length. $t_i$ represents the $ith$ tensor.

First, \textit{Liveness Analysis} reduces the baseline $peak_m$ to $\sum_{i=1}^{N} l_i^f + l_{N}^{b}$ by recycling free tensors amid back-propagation, demonstrating up to $50\%$ of the memory saving. This technique is guaranteed to work on various non-linear architectures, and it is constructed in $\mathcal{O}(N^2)$. \textit{Liveness Analysis} involves high-frequent memory operations on the large chunk memory, while native memory utilities, e.g. cudaMalloc and cudaFree, incur the nontrivial overhead. We address this issue with a preallocated heap managed by the runtime.

Secondly, \textit{Unified Tensor Pool(UTP)} further reduces $peak_m$ to $\sum_{i=1}^{N} (l_i^f \notin checkpoints) + l_{N}^{b}$, where checkpoints represent the compute-intensive layers such as FC and CONV. UTP provides a consolidated memory abstraction to external memory pools to supply for the training. Instead of using naive on-demand data transfers, it hides communications under computations. While the overlapping opportunity is limited given the fixed amount of computations, UTP further introduces a \textit{Tensor Cache} built on GPU to reduce communications.

Finally, \textit{Cost-Aware Recomputation} reduces $peak_m$ to $max(l_i)$, the minimum at the layer-wise granularity. The method keeps track of memory distributions among checkpoints to minimize the extra computations while ensuring $peak_m \leq max(l_i)$.

\subsection{Prerequisites}

\begin{figure}[h]
\centering
\includegraphics[width=0.55\columnwidth]{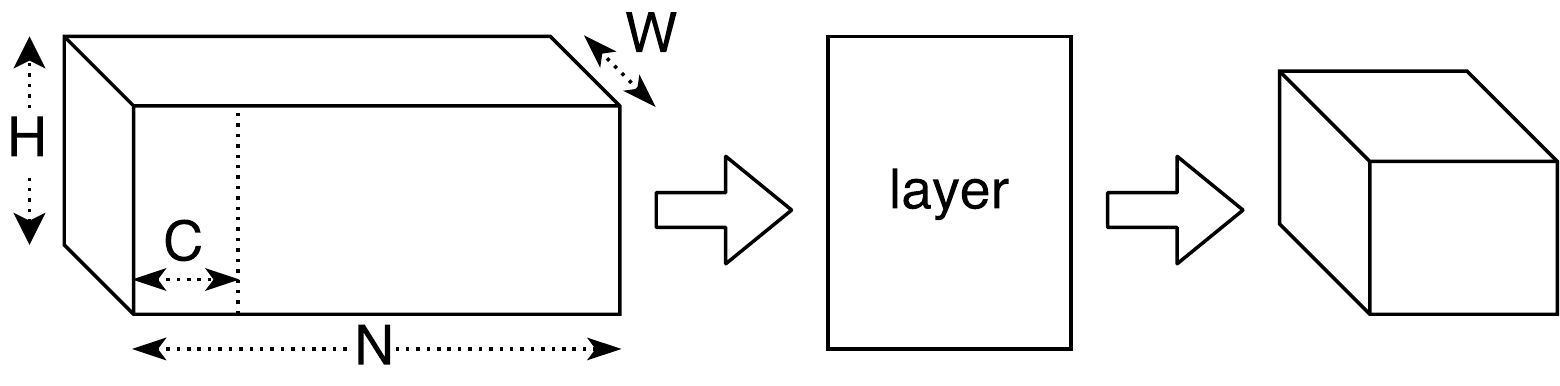}
\caption{The structure of tensors used in DNN. }
\label{tensor}
\end{figure}

\begin{figure*}[!t]
\centering
\includegraphics[width=2.0\columnwidth]{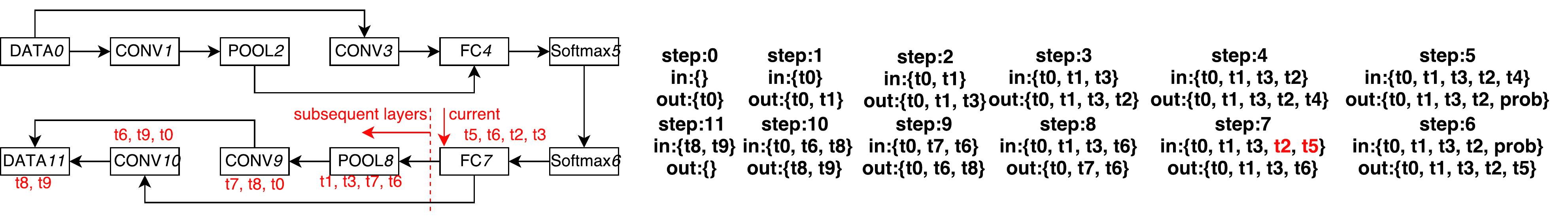}
\caption{Applying \textit{Liveness Analysis} on the nonlinear network shown in Fig.\ref{fan_dep}. The number after the layer name (e.g., DATA0, CONV1, etc.) represents the step, which are calculated by Alg. \ref{compute_route}. We mark the prerequisite tensors for a layer in red, such that $t_7,t_8,t_0$ are required by CONV9. Each $in$ and $out$ set tracks live tensors before and after the layer's computations. We can free $t2$ and $t5$ at step 7 since no subsequent dependencies from POOL8, CONV9, CONV10, and DATA11.}
\label{liveness}
\end{figure*}

A typical DNN network layer computes on a 4-dimension tensor indexed by batches (N), 
image channels (C), height (H) and width (W) (Fig.\ref{tensor}). 
Since cuDNN operates at the layer granularity, we use tensors as the basic memory scheduling unit. 

Alg.\ref{compute_route} describes how SuperNeurons constructs execution steps for nonlinear neural architectures. 
The input is the first network layer; then Alg.\ref{compute_route} recursively explores the subsequent layers in Depth-First Searching (DFS), except that it reaches a join where all prior layers must finish before proceeding. The behavior is achieved by the counter in each layer that tracks the input dependencies (line 5 $\rightarrow$ 6 in Alg.\ref{compute_route}).

\begin{figure}[!t]
\centering
\includegraphics[width=0.75\columnwidth]{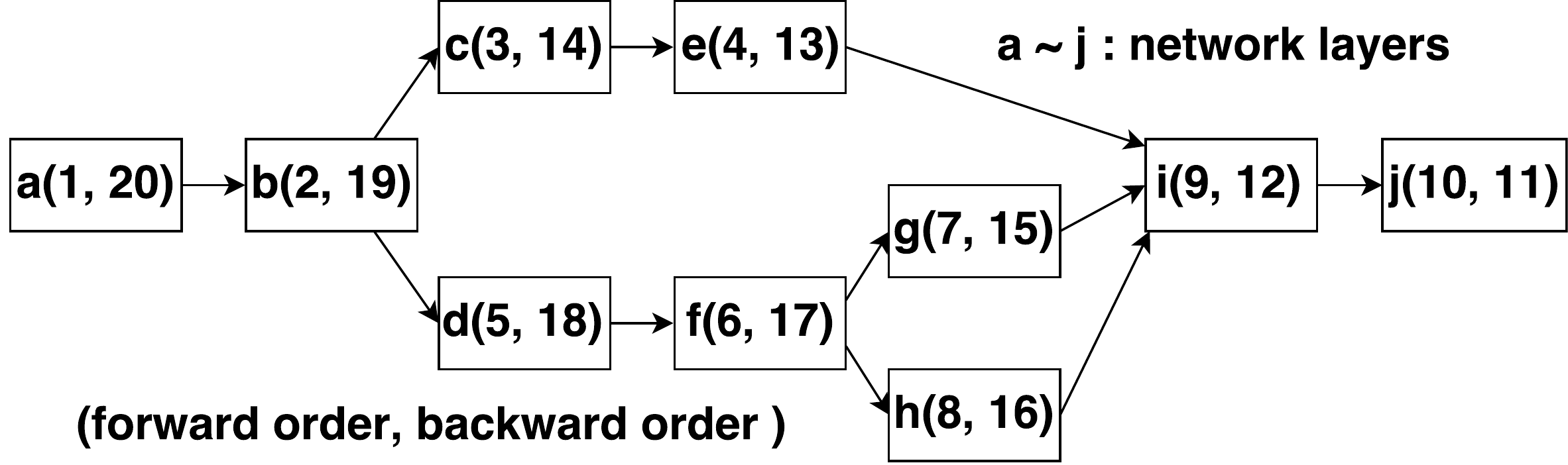}
\caption{Execution route created by Algorithm \ref{compute_route} on a nonlinear network. The left digit represents the forward step, while 
the right digit represents the backward step. }
\label{computation_route}
\end{figure}

Fig.\ref{computation_route} demonstrates an example execution route for a nonlinear network constructed by Alg.\ref{compute_route}.
Each box represents a network layer indexed from $\mathbf{a}$ to 
$\mathbf{j}$. Note that this network has two fan structures
(layer $\mathbf{b}, \mathbf{c}, \mathbf{d}$ and layer $\mathbf{f}, \mathbf{g}, \mathbf{h}$) 
nested together. Alg.\ref{compute_route} successfully identifies layers $\mathbf{e}, \mathbf{g}$ and $\mathbf{h}$
as the prerequisites for executing $\mathbf{i}$.

\begin{algorithm}[!t]
 \caption{Construct execution steps for nonlinear neural architectures}
 \label{compute_route}
 \footnotesize
 \KwData{neural architecture definitions}
 \KwResult{ execution order }
  \Fn(){\construct{layer}}{
     \If{layer is NULL} {
         return
      }
     $layer \rightarrow counter\_inc()$\;
     \If{layer$\rightarrow$get\_counter $<$ size of prev layers } {
         return
      }
      $computation\_route.push(layer)$;\\
	  $next\_layers = b \rightarrow get\_next()$; \\
      \For{$next\_l \in next\_layers$}{
            \construct{next\_l};
      }
	  $reset \ layer\rightarrow counter \ to \ 0$
 }
\end{algorithm}

%\subsection{Construct the Execution Route of Nonlinear Neural Networks}

%\begin{figure}[!t]
%\centering
%\includegraphics[width=0.75\columnwidth]{figure/gpu_malloc}
%\caption{performance deterioration using cudaMalloc and cudaFree.}
%\label{gpu_malloc}
%\end{figure}

%Traditional iterative data-flow analysis takes the worst case of $\mathcal{O}({N^4})$ as
%regular programs involve with conditional branches or loops. The tensor flow of neural networks 
%is static once the architecture is defined. Therefore,

\subsection{ Liveness Analysis and Its Related Issues}

Liveness analysis enables different tensors to reuse the same physical memory at different time partitions. Our runtime implements a simple yet effective variant of the traditional data flow analysis constructed in $\mathcal{O}({N^2})$ for various nonlinear neural networks.
The general procedures are as follows: 
\begin{enumerate}[leftmargin=*]
  \item We construct an $in$ and $out$ set for every layers to track the live tensors before and after the layer, which cost $\mathcal{O}({N})$, where $N$ is the network length.
  \item The runtime populates a layer's $in$ and $out$ sets by checking the dependencies of subsequent layers. It eliminates tensors in $in$ from $out$ if no subsequent layers need them. The cost is $\frac{N(N-1)}{2} \sim \mathcal{O}({N^2})$ as each check costs $N-1$, $N-2$, $...$, $2$, $1$, respectively.
\end{enumerate} 
Fig.\ref{liveness} demonstrates the detailed procedures of \textit{Liveness Analysis} on the network shown in Fig.\ref{fan_dep}. It explicitly lists the content of $in$ and $out$ sets at each steps. For instance, for FC7, $in = {\mathbf{t0}, \mathbf{t1}, \mathbf{t3}, \mathbf{t2}, \mathbf{t5}}$. It needs to create tensor $\mathbf{t6}$ to finalize the current computation. Since $\mathbf{t2}$ and $\mathbf{t5}$ are no longer needed after FC7, runtime eliminates them from FC7's $out$ set (step:7).

\textit{Liveness Analysis} reduces the baseline $peak_m = \sum_{i=1}^{N} l_i^f + \sum_{i=1}^{N} l_i^b$ to $\sum_{i=1}^{N} l_i^f + l_{N}^{b}$. In order to simplify the analysis, let's assume identical memory usages on every layers, i.e. $l_{i}^{f} = l_{i}^{b}$ where $i \in [1, N]$. In the network training, the results of forward pass are needed by the backward propagation\footnote{Not all layers require the previous forward output for the back-propagation, again we simplify the case for the analysis.} \cite{wang2017accelerating, chetlur2014cudnn}. Therefore, the forward total memory usages at step $k$ is $cost^{f}_{k} = \sum_{i=1}^{k} l_i^f$, where $k \leq N$. During the back-propagation, \textit{Liveness Analysis} frees
$l_{i}^f$ and $l_{i}^b$ where $i \in [k+1, N]$ at the backward step $k$ since no future dependencies on them as demonstrated in Fig.\ref{liveness}.
Therefore, the backward total memory usages at step $k$ is $cost^{b}_{k} =\sum_{i=1}^{k} l_i^f + l_{k}^{b}$ and $k \leq N$. Since $l_i > 0$, the $peak_m$ is $max(max(cost^{f}_{k}), max(cost^{b}_{k})) = \sum_{i=1}^{N} l_i^f + l_{N}^{b}$. Therefore, \textit{Liveness Analysis}  saves up to 50\% memory from the baseline.

\begin{figure}[!t]
\centering
\includegraphics[height=0.95in]{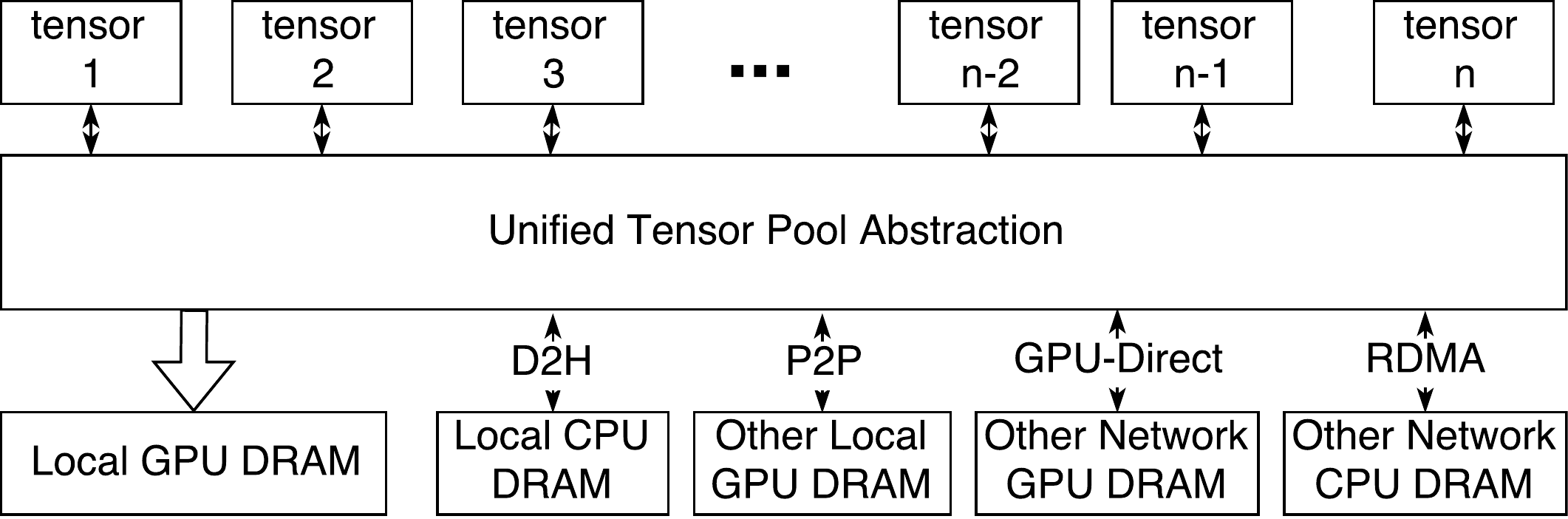}
\caption{The unified tensor pool provides a consolidated memory abstraction to include various physical
memory pools for tensor allocations. }
\label{external_buffer}
\vspace{-0.5 cm}
\end{figure}

\subsubsection{Toward a High Performance Liveness Analysis}
Both the empty initial $in$ set at step 0 and the empty final $out$ set at step 11 in Fig.\ref{liveness} demonstrates \textit{Liveness Analysis} frequently stashes and frees tensors on the fly in a training iteration, while a typical training phase consists of millions of iterations and such intense memory operations incur nontrivial overhead if using the native \emph{cudaMalloc} and \emph{cudaFree} \cite{wang2016blasx}. According to the experiment,
ResNet50 wastes $36.28\%$ of the training time on memory allocations/deallocations with \emph{cudaMalloc} and \emph{cudaFree}. 
To alleviate this performance issue, we implement a fast 
heap-based GPU memory pool utility. The core concept is to remove the allocation/deallocation 
overhead by preallocating a big chunk of GPU memory as a shared memory pool.
Then we divide the entire GPU memory pool into 1KB blocks as the basic storage 
unit. The memory pool contains a list of allocated and empty memory nodes. Each node in 
the two lists contains memory address, occupied blocks and node ID. For an allocation 
request, the memory pool finds the first node with enough free memory from the empty list. 
After that, it updates the empty list and creates a new node in the allocated list to 
track the current allocation.
For a deallocation request, the memory pool locates the node in the
allocated list with the ID-to-node hash-table, then the pool places the node back to the 
empty list.

\subsection{Unified Tensor Pool(UTP) and Its Related Issues}

 If the depth of a neural network goes to $10^3$, the ImageNet training still consumes at least 
 $10^2$GB memory. Therefore, \textit{Liveness Analysis} alone is inadequate for the emerging deep nonlinear
 neural architectures. We provide \emph{Unified Tensor Pool (UTP)} to further alleviate the GPU DRAM shortage by
 asynchronously transferring tensors in/out the external memory.
 \textit{UTP} is a consolidated memory pool abstraction for tensor allocations/deallocations, 
 using various external physical memory such as CPU DRAM, DRAM of other GPUs, or remote CPU/GPU DRAM. 
 In this paper, we focus on the scenario of using local CPU DRAM as an external pool for the fast and efficient interconnect, 
 but the abstraction also applies to other cases shown in Fig.\ref{external_buffer}.
 \textit{UTP} intelligently manages the tensor placement, movement, allocation and deallocation, so that the 
 underlying memory management is entirely transparent to DL practitioners.

\begin{figure}[t]
\vspace{-0.2cm}
\subfloat[][breakdown of execution time by layer types]{\includegraphics[width=1.0\columnwidth]{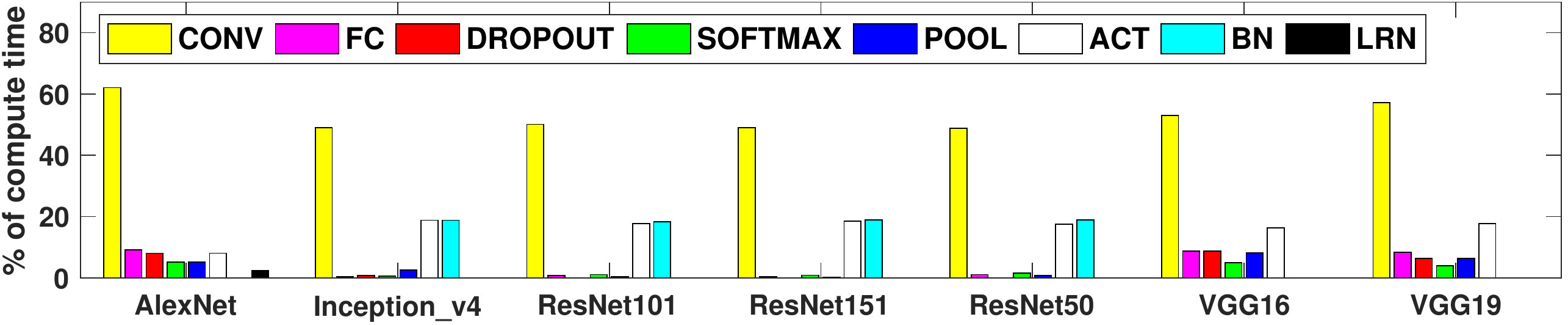}\label{compute_per}} \quad
\subfloat[][breakdown of memory usages by layer types]{\includegraphics[width=1.0\columnwidth]{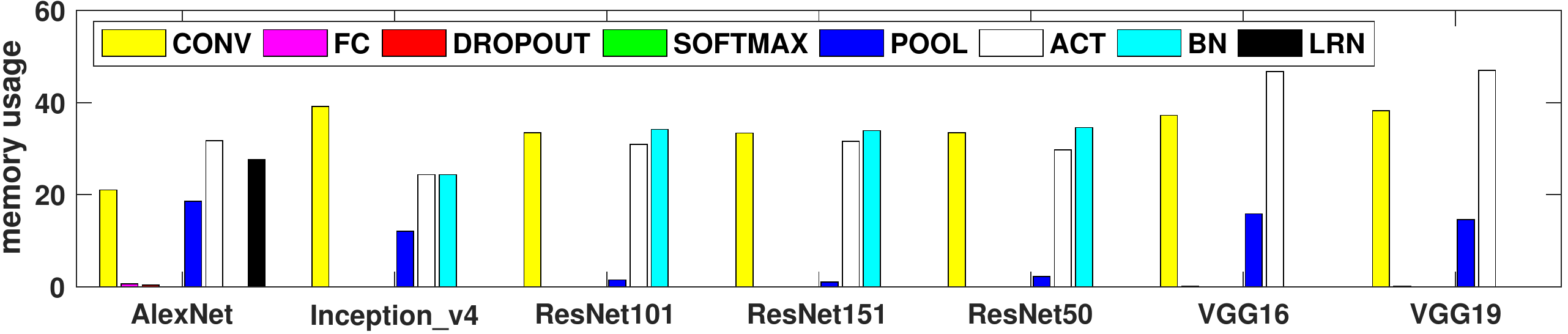}\label{mem_2}} \quad
\caption{The percentages of execution time and memory usages by layer types in different networks. Note that the execution time includes both forward and backward passes.}
\label{breakdown}
\vspace{-0.5cm}
\end{figure}

\begin{figure*}[t]
\vspace{-0.5cm}
\subfloat[][Speed-Centric Recomputation]{\includegraphics[height=0.6in]{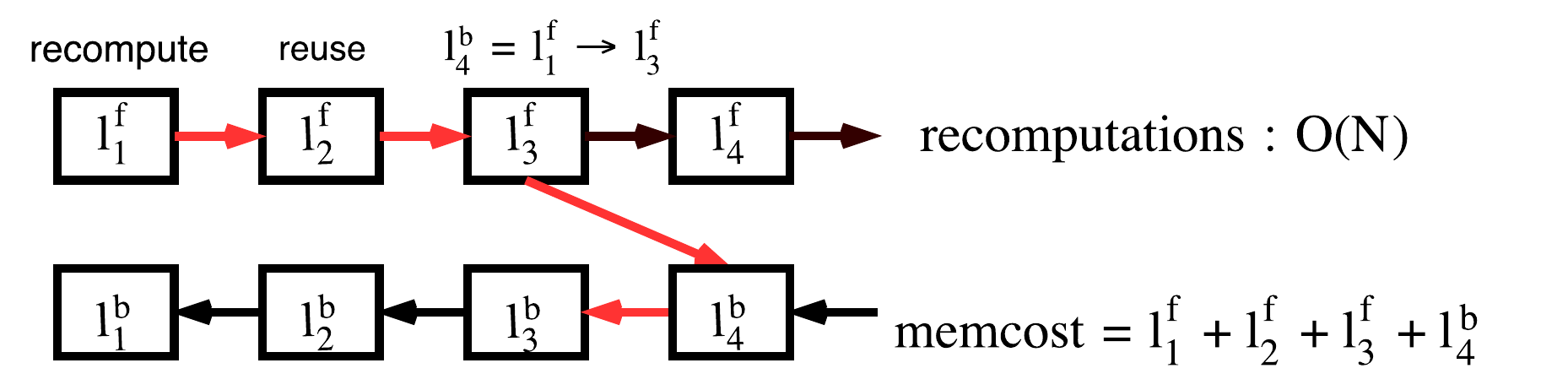}\label{fastest_recomputations}} \quad
\subfloat[][Memory-Centric Recomputation]{\includegraphics[height=0.6in]{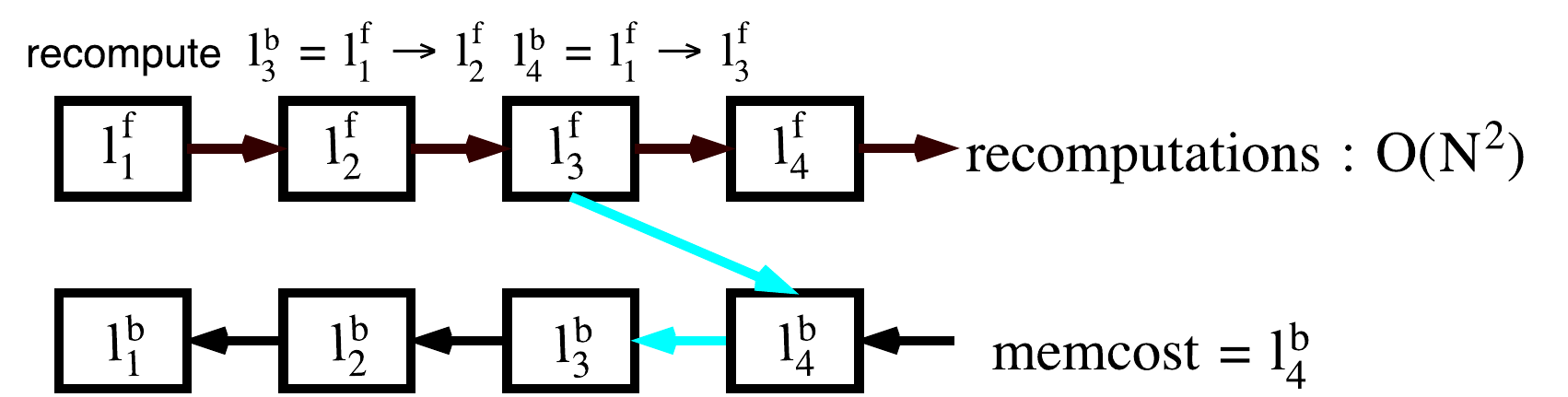}\label{memory_recomputations}} \quad
\subfloat[][Cost-Aware Recomputation]{\includegraphics[height=0.6in]{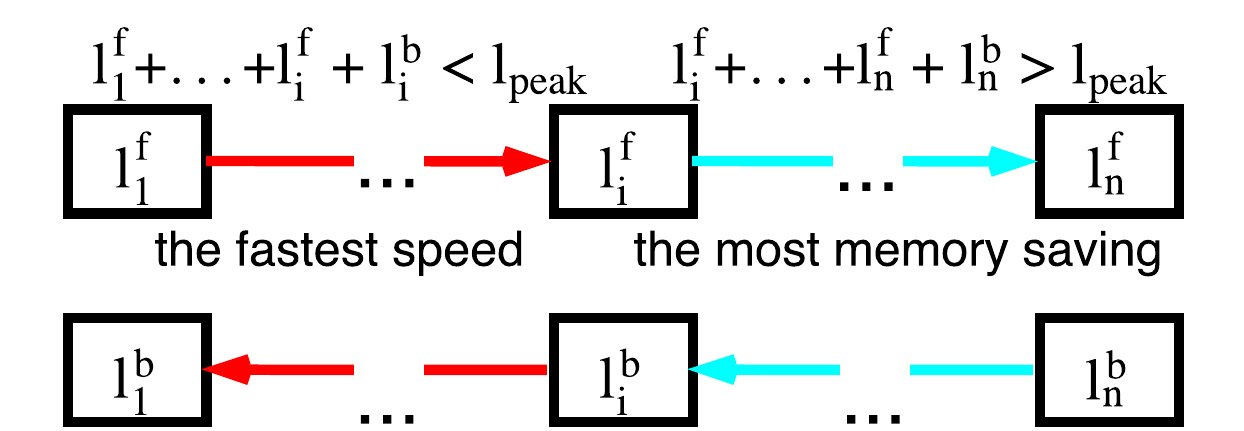}\label{cost_aware_recomputations}} \quad

\caption{The speed-centric strategy only recomputes the segment once, and other backward layers within the segment will reuse the recomputed tensors. Thus, it only incurs $\mathcal{O}(N)$ additional computations, but $memcost$ is $\sum_{i=1}^{seg}l_{i}^{f} + l_{seg}^{b}$. The memory-centric strategy recomputes forward dependencies every time for each backward layers. Though it incurs $\mathcal{O}(N^2)$ additional computations, $memcost$ is the least, i.e. $l_{i}^{b}$. Cost-Aware Recomputation profiles the memory usages across recomputation segments. It uses the speed-centric strategy (red) if $memcost$ of a segment is less than $l_{peak}$, and the most memory saving strategy (blue) otherwise.}
\label{recomputation_strategies}
\end{figure*}

\subsubsection{Basic UTP Memory Management: Memory Offloading and Prefetching}
Not all the layers are suitable for \textit{Offloading} and \textit{Prefetching}. 
We define transferring tensors from GPU to external physical pools as \emph{Offloading}, 
and the reversed operation as \emph{Prefetching}. Fig.\ref{compute_per} 
and Fig.\ref{mem_2} demonstrate that POOL, ACT, BN and LRN all together occupy 
over $50\%$ of the total memory, while their computations only account for an average of 
$20\%$ of the entire workload. Thus, offloading these layers incurs a great overhead 
due to the insufficient overlapping of communications and computations. 
It is also not fruitful to offload on Dropout, Softmax and FC layers 
since they only use less than 1\% of the total memory. Therefore, we only offload the tensors from CONV layers.

\textbf{\textit{Offloading}}:the runtime asynchronously transfers the forward outputs of CONV layers 
to the preallocated pinned CPU memory. It records an event for this data transfer and frees the tensor's GPU 
memory once the event is completed. The runtime has an independent thread running in the background to check events of memory copies;
and this enables GPU-to-CPU data transfers to overlap with the forward computations starting from the current CONV layer to the next one.

\textbf{\textit{Prefetching}}:the runtime asynchronously brings the offloaded and soon to be reused 
tensors back to the GPU DRAM. At any CONV layers in the backward, the runtime asynchronously fetches 
the required tensors for the previous CONV layer. This enables the CPU-to-GPU data transfer to overlap 
with the backward computation starting from the current CONV layer to the previous one.

\textit{Offloading} and \textit{Prefetching} reduce $peak_m$ after \textit{Liveness Analysis} to 
$\sum_{i=1}^{N} (l_i^f \notin checkpoints) + l_{N}^{b}$, where $checkpoints = \{CONV\}$. Since layers in 
$checkpoints$ are offloaded, the total memory consumption at each backward steps is $cost(k) = \sum_{i=1}^{k} (l_i^f \notin checkpoints) + l_{k}^{b}$,
where $k \in [1, N]$. The memory usage of each layers is non-negative, thus $peak_m = max(cost(k))$ is $\sum_{i=1}^{N} (l_i^f \notin checkpoints) + l_{N}^{b}$.

\begin{algorithm}[t]
%\scriptsize
\footnotesize
\caption{The basic LRU operations}
\label{LRU_alg}
\DontPrintSemicolon
\KwData{Tensor ($T$) and $LRU$}
\KwResult{Tensor with the GPU memory.}

\Fn{$LRU.in$ ($T$)} {
    $T.Lock  \leftarrow false$ \tcc*{A layer will lock its dependent tensors in the computation.}
	$LRU.insertFront(T)$
}

\Fn{$LRU.out$ ($T$)}{
    $freedMem \leftarrow 0$\\
    \While{$freedMem < T.size$} {
		$T^{\prime} = LRU.getLastUnlockedTensor() $ \\
		$freedMem = freedMem + T^{\prime}.size$ \\
		$remove ~ T^{\prime} ~ from ~ LRU ~ list $   \\
		\textit{offload} $ T^{\prime}.GA$  to $T^{\prime}.CA $ \tcc*{CA is CPU Addr}
    }
	$T.GA \leftarrow Malloc(T.size) $
}

\Fn{$Check$ ($LRU$, $T$)}{
    $isFound \leftarrow LRU.find(T)$\\
    \If{$isFound = false$} {
        $ T.GA \leftarrow Malloc(T.size)$ \tcc*{GA is GPU Addr}
        \If{$T.GA = \emptyset$} { 
            $T.GA \leftarrow LRU.out()$\\
        }
        $LRU.in(T)$ \tcc*{cache miss}
    } \Else { 
        % \tcc*{the Most Frequently Used is at the list front.}
        $ LRU.placeToFront(T)$    \tcc*{cache hit}
    }
    \Return $T.GA$ 
}

\end{algorithm}

% \begin{figure}[!t]
% \centering
% \includegraphics[height=0.75in]{figure/tensor_state}
% \caption{State transitions of tensors in UTP. }
% \label{tensor_state}
% \vspace{-0.5cm}
% \end{figure}
%
% \subsubsection{State Transition of Tensors }
% The underlying data of a tensor can be either placed on CPU or GPU DRAM. The runtime utilizes 4 states
% to identify the tensor status: GPU, CPU, GPU2CPU and CPU2GPU, shown in Figure \ref{tensor_state}. GPU and CPU
% mark the actual memory locations at a particular moment, and the other two (top and bottom) mark the two asynchronous
% GPU2CPU and CPU2GPU data transfer in Memory Offloading and Prefetching, respectively. GPU2CPU is necessary
% because a GPU will reuse a tensor while it's asynchronously transferring the data to CPU if
% the reuse distance is small. In this case, the runtime directly sets the tensor state to GPU without additional memory
% operations. A tensor changes its state from CPU to CPU2GPU if asynchronous prefetching issued,
% and the runtime automatically sets the tensor's state to GPU once the CPU2GPU event is completed.
% GPU is also able to exchange the data on CPU DRAM in a synchronous fashion.

\subsubsection{Caching Tensors on GPU DRAM }
While the overlapping opportunity is limited given the fixed amount of computations in an iteration,
the aforementioned on-demand \textit{Prefetching}/\textit{Offloading} protocol can quickly exhaust the chance.
Nowadays CPU-to-GPU data movements over PCI-E, GPU-to-GPU data movements over the same PCI-E switch, 
and GPU-to-remote GPU over GPU-Direct RDMA deliver a practical speed
of 8 GB/s, 10 GB/s, and 6 GB/s, but transferring Gigabytes data in each training iterations incurs the nontrivial 
overhead. Therefore, this on-demand tensor transfer protocol must be optimized.
SuperNeurons proposes a \textit{Tensor Cache} to exploit the temporal localities of tensors. It caches tensors on GPU DRAM 
to maximize their reuses and to minimize the global communications.
With \textit{Prefetching} and \textit{Offloading}, the runtime only triggers data transfers
when GPU DRAM is insufficient.

We adopt Least Recent Used (LRU) tensor replacement policy to build \textit{Tensor Cache}. Since the back-propagation demonstrates 
the head-to-tail and tail-to-head computation pattern, it subjects the most recent used tensors to the earliest
reusing as suggested in Fig.\ref{liveness}. This motivates us to design \textit{Tensor Cache} with a simple variant of LRU. 
While there are other sophisticated cache replacement policies might be better fit the scenario, thorough discussions of them 
fall out the scope of this paper.

Alg.\ref{LRU_alg} demonstrates the three key operations of proposed LRU. 1) \textit{LRU.in} function intends to place a tensor into LRU.
Each tensor has a lock, and a tensor cannot be removed from LRU if locked. A layer will lock dependent tensors at calculations. 
LRU is implemented by a list with the front as Most Frequently Used (MFU) and the tail otherwise. 2) \textit{LRU.out} function intends to remove
enough bytes for a new tensor. It offloads the unlocked Least Recent Used tensors to CPU RAM till having enough free memory for the new one.
3) \textit{Check} function decides what to operate on the tensor. It takes in a tensor to check if the tensor is in $LRU$ based on the object address (line 2). If found, we place the tensor to the MFU position, i.e. the list front (line 9), and return the tensor's 
GPU address. This is the hit scenario. If not found, we call \textit{LRU.out} to free enough memory for the new tensor before inserting it into LRU, and this is the miss scenario.

% \begin{figure}[!t]
% \centering
% \includegraphics[width=0.55\columnwidth]{figure/recomputation_demo}
% \caption{ The concept of re-computation for memory. }
% \label{recomputation_demo}
% \vspace{-0.5cm}
% \end{figure}

\subsection{Cost-Aware Recomputation}
\label{recomputations}
POOL, ACT, LRN and BN all together use an average of $50\%$ memory, 
while their forward computations only account for less than $10\%$ of the total time. 
This exposes additional $50\%$ memory savings with a fraction of performance 
loss by recomputing the forward dependencies in the back-propagation. 
Basically, the runtime frees the tensors in cheap-to-compute layers such as 
POOL for reconstructions. In general, there are memory-centric and speed-centric 
strategies for the recomputation for memory.

The speed-centric strategy keeps the recomputed tensors so that other backward layers 
can directly reuse them. Fig.\ref{fastest_recomputations} denotes the procedures in red.
At the backward step on $l_4^{b}$, it performs a forward pass from $l_1^{f}$ to $l_3^{f}$ 
to get dependencies for $l_4^{b}$. It keeps $l_1^{f}, l_2^{f}$ so that they can be re-used 
for the backward computation on $l_{3}^{b}$ and $l_{2}^{b}$.
% \textcolor{red} {it is hard to match what you said to Figure 9. what are color arrow means? you said
% l1f l2f feed l3b and l2b, but I can not see them in the figure. I suggest you to put data depedency graph here and mark which are recomputed. similar for figure b and c}
MXNet \cite{chen2016training} adopts this strategy. It incurs the least $\mathcal{O}(N)$ additional computations, 
but $memcost$ is $\sum_{i=1}^{seg}(l_i^f) + l_{seg}^{b}$. $memcost$ will exceed $l_{peak}$ if $l_{peak}$ is within the segment.

The memory-centric strategy always recomputes the dependencies for each backward layer. 
In contrast to the speed-centric one, it fully exploits the memory-saving opportunity 
by freeing the recomputed intermediate results. For example, it recomputes 
$l_1^{f} \rightarrow l_3^{f}$ for $l_4^{b}$, while it recomputes $l_1^{f} \rightarrow l_2^{f}$ again for $l_3^{b}$ as demonstrated by the blue lines in Fig.\ref{memory_recomputations}. The $memcost$ stays at $l_{i}^{b}$ guaranteed to be $ \leq l_{peak}$, but the strategy incurs $\mathcal{O}(N^2)$ additional computations.

We present a new \textit{Cost-Aware Recomputation} that leverages the advantages of both methods.
It is motivated by the observation that the memory costs of most recomputation segments are 
$\leq l_{peak}$, i.e. $\sum_{i=1}^{seg}(l_i^f) + l_{seg}^{b} \leq l_{peak}$. That implies
we can leverage the least recomputations in the speed-centric strategy while still guarantees 
the memory usage to be $\leq l_{peak}$ as in the memory-centric strategy. The general procedures of 
\textit{Cost-Aware Recomputation} are as follows:
\begin{enumerate}[leftmargin=*]
  \item the runtime iterates over all the layers to find $l_{peak} = max(l_{i})$ as the threshold.
  \item In a recomputation segment, the runtime applies the speed-centric strategy (marked by red in Fig.\ref{cost_aware_recomputations} ) if $\sum_{i=1}^{seg}(l_i^f) + l_{seg}^{b} \leq l_{peak}$, and the memory-centric strategy (marked by blue in Fig.\ref{cost_aware_recomputations}) otherwise.
\end{enumerate}
Table.\ref{cost_aware} summarizes the extra recomputation for two basic strategies and \textit{Cost-Aware Recomputation}. 
Our cost-aware method ensures $peak_m$ to be consistent with the memory-centric strategy, while the extra recomputations are comparable to the speed-centric strategy.

\textit{Cost-Aware Recomputation} finally reduces $peak_m$ to $max(l_i)$. Previously, \textit{Liveness Analysis} and \textit{Offloading} jointly reduce the $cost^{b}_{k}$ to $\sum_{i=1}^{k} (l_i^f \notin checkpoints) + l_{k}^{b}$. Since \textit{non-checkpoints} layers will be freed for recomputations, only the nearest \textit{checkpoint} layer exists in the GPU memory. Thus, $cost^{b}_{k} = l_{checkpoint}$. During the recomputations, $cost^{b}_{k}$ can be either $\sum_{i=1}^{k}(l_i^f) + l_{k}^{b} \leq l_{peak}$ or $l_{i}^b$ depending what recomputation strategies to use. Whereas, \textit{Cost-Aware Recomputation} guarantees $cost^{b}_{k} \leq l_{peak} = max(l_i)$ (see analyses above). Thus, the final network wide $peak_m = max(cost^{b}_{k}) = l_{peak}$, which is the minimal $peak_m$ achievable at the layerwise granularity.

\begin{table}[!t]
\scriptsize
\setlength{\tabcolsep}{0.25em}
\caption{ The counts of recomputations (extra) and $peak_m$ using the speed-centric, the memory-centric and Cost-Aware Recomputation. }
\label{cost_aware}
\begin{tabular}{c c c c c c c c}
  \hline
                              &   \multicolumn{2}{c}{\textbf{speed-centric}}  &  \multicolumn{2}{c}{\textbf{memory-centric}} &  \multicolumn{2}{c}{\textbf{cost-aware}} &  \\   \hline
                              &  extra     &  $peak_m$                &       extra     &  $peak_m$              &       extra     &  $peak_m$      &  \\
           AlexNet            &  14        &  993.018                   &       23        &  886.23                  &  17         &  886.23               &  \\
           ResNet50           &  84        &  455.125                   &       118       &  401                     &  85         &  401                  &  \\
           ResNet101          &  169       &  455.125                   &       237       &  401                     &  170        &  401                  &  \\
   \hline
\end{tabular}
\vspace{-0.15in}
\end{table}

%by taking advantage of both
%methods. 
%
%max backward layer 2725MiB stash with liveness analysis

%backward conv_2: 2271 MB
%                 2623 MiB
%                 2555 MB
%        
%without free:     the segment needs 352 MB for forward computations
%with free: 	      the segment needs 284 MB for forward computations 
%difference is at the pool -> conv2, which is 68.34375
%Bottleneck layer: 1134MiB, pool_1, 

%Memory costs of network layers drastically vary  
%
%When layer i+1, i+2 and i+3 finish the 
%forward computations, the runtime frees their corresponding output tensors $\mathbf{t1}$, $\mathbf{t2}$, and $\mathbf{t3}$.
%In the backward pass, the runtime reconstructs $\mathbf{t3}$ by doing another forward pass in
%the segment from the checkpoint layer to layer i+2. Since the tensors after layer i+3 are freed
%by the liveness analysis, this saves $\mathbf{t2}$ and $\mathbf{t3}$. 

\subsection{Finding the Best Convolution Algorithm under the Memory Constraint}
\label{workspace}

The speed of CONV layers significantly impacts the training as it accounts for over $50\%$ of total computing time (Fig.\ref{breakdown}). cuDNN provides several convolution algorithms, e.g. using FFT, Winograd and GEMM, for different contexts. Some of them, FFT in particular, require temporary convolution workspaces to delivery the maximal speed as demonstrated in Fig.\ref{figure1}. Therefore, the memory is also a critical factor to the high-performance training.

We implement a dynamic strategy for allocating convolution workspaces. It is dynamic because the 
memory left for convolution workspaces constantly changes in every steps according to 
\textit{Liveness Analysis}, \textit{UTP} and \textit{Cost-Aware Recomputation}. 
Since convolution workspaces do not affect the functionality, the allocations of functional tensors 
such as data and parameters are prioritized. Then the runtime steps into each layer to profile 
free bytes left in GPU DRAM after those memory techniques being applied. With free bytes information
at individual layers, the runtime benchmarks all the memory-feasible convolution algorithms to
pick up the fastest one. Please note the runtime skips convolution algorithms that require more 
memory than it can provide. Each layer selects the fastest algorithm under
the remaining GPU DRAM, and therefore maximize the performance of CONV layers and the entire training.

\begin{figure*}[t]
\subfloat[][liveness analysis]{\includegraphics[width=0.7\columnwidth]{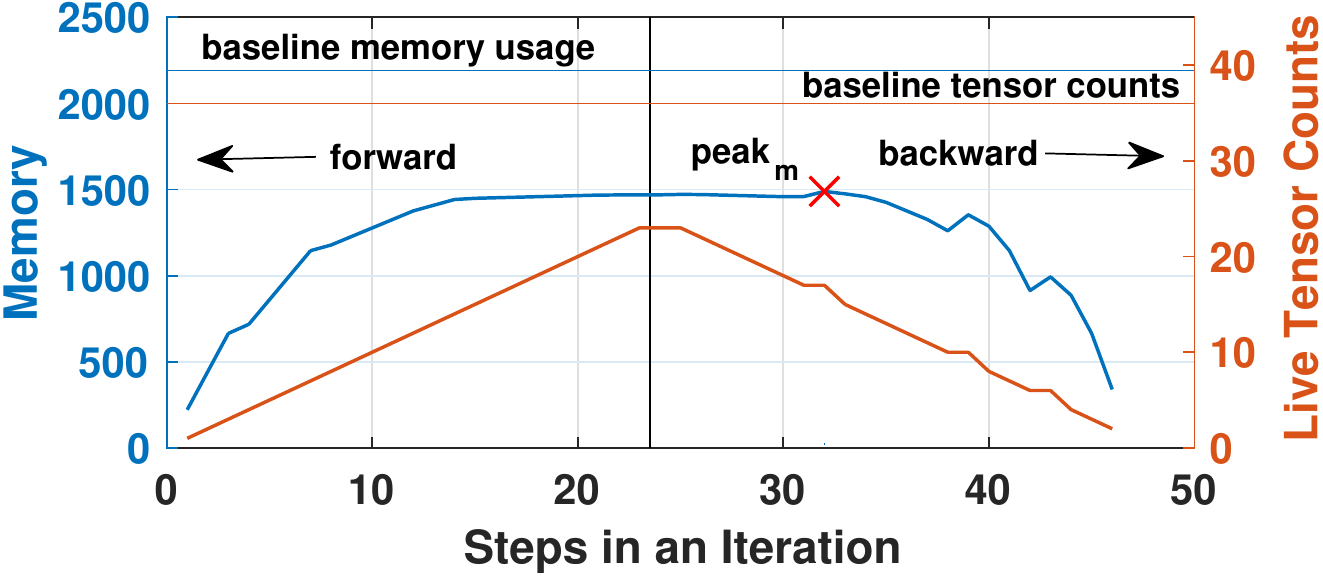}\label{liveness_eval}} 
\subfloat[][prefetching/offloading + liveness]{\includegraphics[width=0.7\columnwidth]{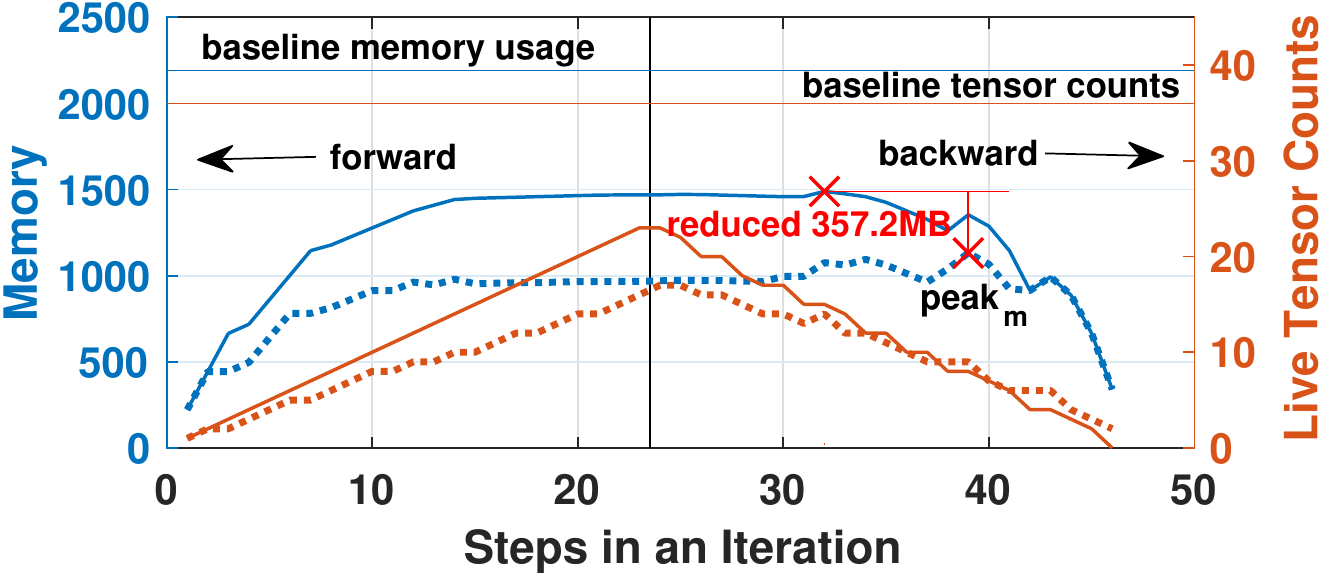}\label{prefetch_eval}}
\subfloat[][recomputation + previous two]{\includegraphics[width=0.7\columnwidth]{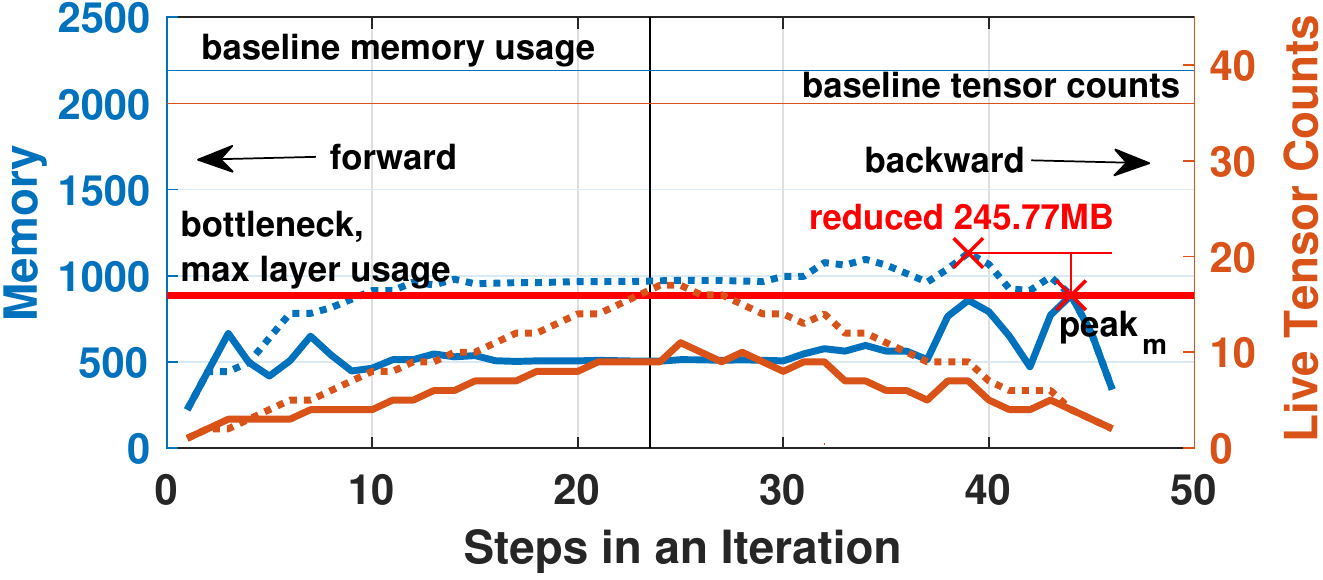}\label{recomputation_eval}}

\caption{The evaluations of \textit{Liveness Analysis}, \textit{Prefetching/Offloading} and \textit{Cost-Aware Recomputation} on AlexNet at the batch size of 200. AlexNet has 23 layers, and a training iteration consists of 1 $\rightarrow$ 23 forward steps and 24 $\rightarrow$ 46 backward steps. The blue curve (left axis) depicts memory usages at each step, while the orange curve (right axis) depicts live tensor counts at each step. 
(a) demonstrates how \textit{Liveness Analysis} affects memory usages w.r.t the baseline (horizontal lines). (b) demonstrates how \textit{Offloading/Prefetching} improve \textit{Liveness Analysis} by comparing the memory usages of both techniques (blue dashed lines in (b)) with \textit{Liveness} alone (solid blue curve in (b)). Similarly, (c) demonstrates how \textit{Cost-Aware Recomputation} improve the previous two; and dashed lines in (c) are from (b).}
\label{memory_techniques}
\vspace{-0.2cm}
\end{figure*}

\section{Evaluations}
In this section, we present the results of our experimental studies that evaluate each memory and performance techniques in SuperNeurons. 
We also did end-to-end evaluations against TensorFlow, MXNet, Caffe and Torch on various neural networks to justify the design.  

\subsection{Components Evaluations}
\subsubsection{ Memory Optimizations }
We use the naive network-wide tensor allocation strategy as the baseline. Thus, the $peak_m$ of baseline is 
$\sum_{i=1}^{N} l_i^f + \sum_{i=1}^{N} l_i^b$
, where $N$ is the network length. (defined in Sec.\ref{method}).
Since cuDNN operates at the layerwise granularity, $peak_m$ is bounded by the maximal memory usage among layers, i.e. $l_{peak}$.
 
\textbf{\textit{Liveness Analysis}} reduces the baseline's $peak_m$ to $\sum_{i=1}^{N} l_i^f + l_{N}^{b}$. 
Fig.\ref{liveness_eval} demonstrates how \textit{Liveness Analysis} affect memory usages
and live tensor counts at each forward/backward steps on AlexNet.  
\footnote{the structure of AlexNet is
CONV1$\rightarrow$RELU1$\rightarrow$LRN1$\rightarrow$POOL1 \\
$\rightarrow$CONV2$\rightarrow$RELU2
$\rightarrow$LRN2$\rightarrow$POOL2$\rightarrow$CONV3$\rightarrow$RELU3 \\
$\rightarrow$CONV4$\rightarrow$RELU4$\rightarrow$CONV5$\rightarrow$RELU5$\rightarrow$POOL5$\rightarrow$FC1
$\rightarrow$RELU6$\rightarrow$Dropout1$\rightarrow$FC2$\rightarrow$RELU7$\rightarrow$Dropout2
$\rightarrow$FC3$\rightarrow$Softmax }
Since AlexNet has 23 layers, there are 23 forward steps and 23 backward steps. 
The central vertical line separates forward and backward while each of them contains 23 computational
steps. The baseline allocates 36 data tensors consuming 2189.437MB, 
while \textit{Liveness Analysis} uses up to 17 tensors with a peak memory usage of 1489.355MB. 
This demonstrates 31.9\% improvement over the baseline in terms of $peak_m$.
It is also observable that the location of $peak_m$ is not necessarily consistent with 
the peak tensor count. This confirms our claim that the memory are unevenly distributed across
network layers.

To verify the cost model, i.e. $cost^{b}_{k} =\sum_{i=1}^{k} l_i^f + l_{k}^{b}$, 
we delve into the memory usages of peak layer. Fig.\ref{liveness_eval} suggests the 32th step
reaches $peak_m$. This corresponds to the backward POOL5 in AlexNet, and $k = 14$ because of 46 - 32.   
The forward layers that are before and include POOL5 stash 5 tensors, consuming 1409.277MB ($\sum_{i=1}^{14} l_i^f$), 
while the backward POOL5 stashes 3 tensors, consuming 80.078MB ($l_{14}^{b}$). Therefore, $cost^{b}_{14} = $  1409.277 + 
80.078 = 1489.355MB, which is consistent with the measured $peak_m$.

\textbf{\textit{Prefetching and Offloading}} reduces the $peak_m$ after \textit{Liveness Analysis} to 
$\sum_{i=1}^{N} (l_i^f \notin checkpoints) + l_{N}^{b}$. Fig.\ref{prefetch_eval} demonstrates the updated 
memory usages and live tensor counts after \textit{Prefetching/Offloading} being applied on the top of \textit{Liveness Analysis}.
We set CONV layers as checkpoints for offloading.
The new $peak_m$ is 1132.155 MB at the 39th step or POOL2 backward. 
It further reduces 357.2MB on the previous $peak_m$ or
total 48.29\% improvement over the baseline's $peak_m$. The new $peak_m$ shifts from POOL5 to POOL2
because of the number of CONV layers ahead of them. CONV1, CONV2, CONV3, and CONV4 are located before POOL5; and
they consume 221.56MB, 142.38MB, 49.51MB and 49.51MB, respectively, The runtime 
offloads CONV $1\sim4$ to CPU RAM and prefetches CONV5. This leads the new memory usage of POOL5 to be 
1489.355 - 221.56 - 142.38 - 49.51 = 1075.9MB, which is less than the measured new $peak_m$ 1132.155 MB at POOL2.

To verify the updated cost model, i.e. $\sum_{i=1}^{k} (l_i^f \notin checkpoints) + l_{k}^{b}$, we compare the calculated 
live tensor count from the model with the actual measurement. There are 2 checkpoints, CONV1 and CONV2, before POOL2; 
and the runtime prefetches CONV2 in the backward. As a result, the calculated live tensor count at POOL2 is 10 (measured live tensors before POOL2)
- 1 (CONV1) = 9. This is same to our actual measurement of 9 tensors at POOL2. 
Therefore, the updated cost model after \textit{Prefetching/Offloading} is still valid.

Finally, \textit{\textbf{Cost-Aware Recomputation}} reduces $peak_m$ to $\max( l_i )$. 
In theory, $\max( l_i )$ is the minimal $peak_m$ at the layerwise granularity as cuDNN 
needs at least stash the tensors in a layer to compute. Fig.\ref{recomputation_eval} demonstrates
stepwise memory usages and live tensor counts with all three techniques.
We profile that $max(l_i) = 886.385 MB$ at the backward LRN1 by iterating through every layer.
Fig.\ref{recomputation_eval} demonstrates a $peak_m$ of 886 MB at the 44th step, which is the backward of LRN1.
Therefore, three proposed memory saving techniques successfully reduce the $peak_m$ from $\sum_{i=1}^{N} l_i^f + \sum_{i=1}^{N} l_i^b$ to $max(l_i)$.

\subsubsection{ Speed Optimizations }
\label{speed_opt}
The runtime equips with a GPU Memory Pool and a Tensor Cache to improve the performance of memory techniques and a dynamic strategy for allocating convolution workspaces to accelerate the training speed. More specifically, GPU Memory Pool amortizes the non-trivial overhead of high-frequent memory allocations/deallocations in \textit{Liveness Analysis}; and Tensor Cache enables tensor reusing to minimize data transfers in \textit{Prefetching/Offloading}. Fig.\ref{recomputation_eval} demonstrates the GPU free space dynamically changes at each forward and backward step due to 3 memory techniques. The runtime allocates convolution workspaces within the free memory at a step. As a result, the performance is optimized at individual layers under different stepwise memory constraints.

\begin{table}[!t]
\scriptsize
\setlength{\tabcolsep}{0.15em}
\caption{The improvement of GPU memory pool over cudaMalloc and cudaFree on various networks. The batch size for AlexNet is 128,
while the rest is 16.}
\label{gpu_malloc_eval}
\begin{tabular}{c c c c c c c c}
  \hline
    \textbf{img/s}         &   AlexNet    &  VGG16       &  InceptionV4       & ResNet50 & ResNet101  & ResNet152  \\   \hline
    CUDA                       &   359.4      &  12.1        &  6.77              &  21.5    &  11.3      & 7.46       \\
    Ours                   &   401.6      &  14.4        &  10.0              &  32.9    &  18.95     & 13.2       \\
    speedup                &   1.12x      &  1.19x       &  1.48x             &  1.53x   &  1.68x     & 1.77x      \\
   \hline
\end{tabular}
\end{table}
 
\begin{table}[!t]
\scriptsize
\setlength{\tabcolsep}{0.25em}
\caption{Communications with/without Tensor Cache. We benchmark the result on AlexNet by increasing
the batch size from 256 to 1024. }
\label{tensor_cache_eval}
\begin{tabular}{c c c c c c c c}
  \hline
    \textbf{Communications in GB} &  256     &  384       &  512       &  640      &  896       & 1024  \\   \hline
    Without Tensor Cache    &   2.56   &  3.72      &  4.88      &  6.03     &  8.35      & 9.50       \\
    Tensor Cache            &   0      &  0         &  0         &  0        &  0         & 0.88       \\
   \hline
\end{tabular}
\end{table}
 
\textbf{\textit{GPU Memory Pool}} amortizes the non-trivial overhead of intensive memory operations in \textit{Liveness Analysis} by preallocating a big chunk of GPU memory.
Table~\ref{gpu_malloc_eval} illustrates the performance improvement of using \textit{GPU Memory Pool} over cudaMalloc and cudaFree. Linear networks such as AlexNet and VGG involve much fewer memory operations than
nonlinear ones such as InceptionV4 and ResNet$50\rightarrow152$ due to the limited depth. Therefore, the speedups on nonlinear networks (ResNet 50$\rightarrow$152 and InceptionV4) are more significant than linear networks (AlexNet, VGG).

%The deeper a network goes, the more memory operations it has.
%Therefore, the speedups on deep and wide neural networks increases along with the depth in ResNet $50\sim152$.

\begin{figure}[!t]
\centering
\includegraphics[width=0.85\columnwidth]{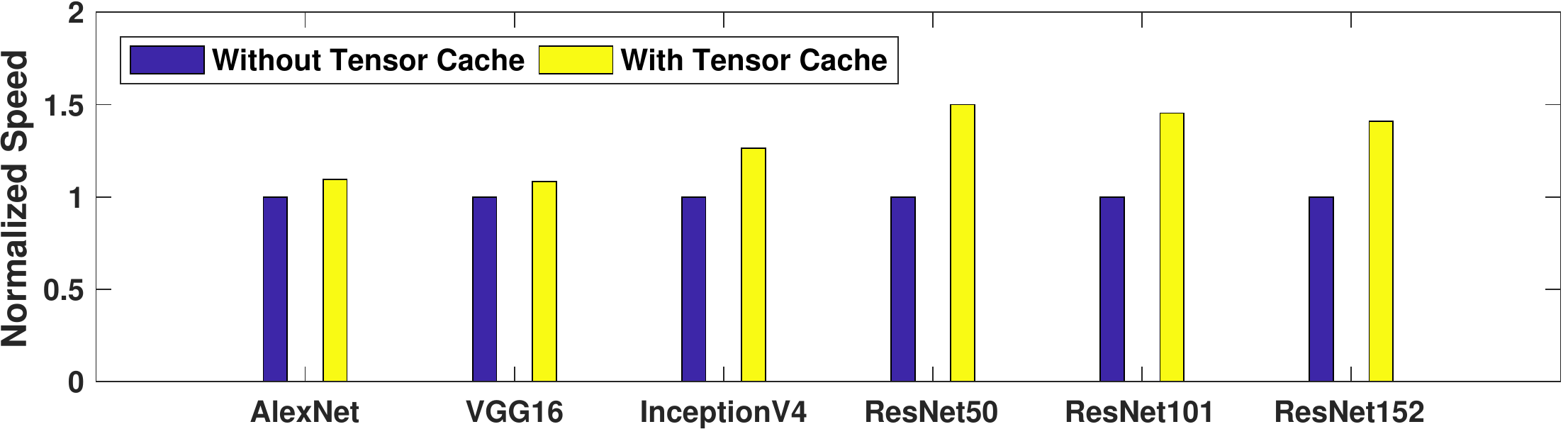}
\caption{ Normalized performance with and without \textit{Tensor Cache}. The batch size of AlexNet is 128, and 32 for the rest. }
\label{tensor_cache_speed}
\end{figure}

\textit{\textbf{Tensor Cache}} intends to reduce unnecessary data transfers in \textit{Prefetching/Offloading}.
Specifically, the offloading is unnecessary if a network can fit into the GPU DRAM. In Table~\ref{tensor_cache_eval}, 
we can see \textit{Tensor Cache} successfully avoids communications at batch sizes of 256 $\rightarrow896$, while the
communications, in the scenario without \textit{Tensor Cache}, linearly increase along batch sizes. 
The training performance will deteriorate if communications outweigh computations. Fig.\ref{tensor_cache_speed} demonstrates up to 33.33\% performance loss without using \textit{Tensor Cache}. It is also noticeable that the speedup on linear networks (AlexNet, VGG16) is less significant than nonlinear ones (ResNet50$\rightarrow$152, Inception). In general, the computation intensity of a linear network layer is far more than the non-linear one.
Because their communications can overlap with computations in \textit{Prefetching/Offloading}, \textit{Tensor Cache} does not provide the comparable speed up for AlexNet and VCG16.

\begin{figure}[t]
\subfloat[][batch=100, pool = 3G]{\includegraphics[width=0.45\columnwidth]{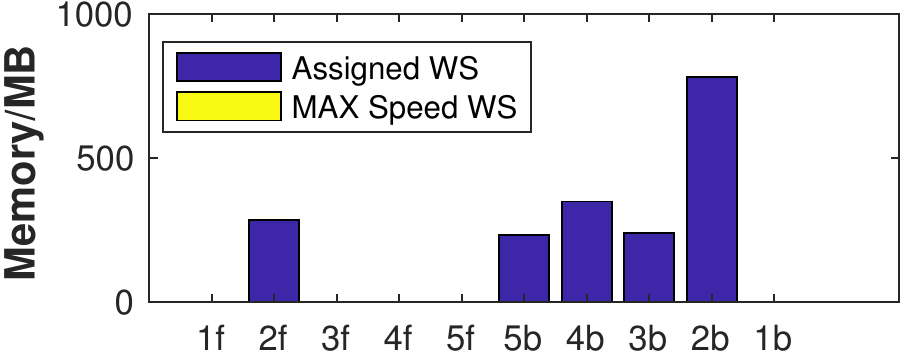}\label{dyna_conv_batch_2}} \quad
\subfloat[][batch=300, pool = 3G]{\includegraphics[width=0.45\columnwidth]{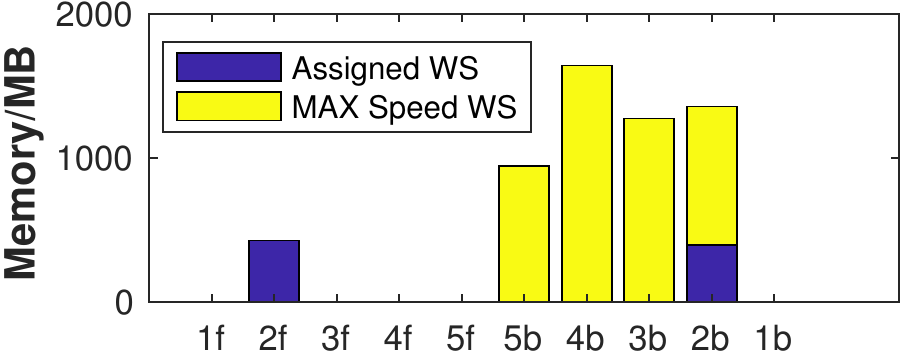}\label{dyna_conv_batch_1}} \quad
 \\
\subfloat[][203(imgs/s), pool = 3G]{\includegraphics[width=0.45\columnwidth]{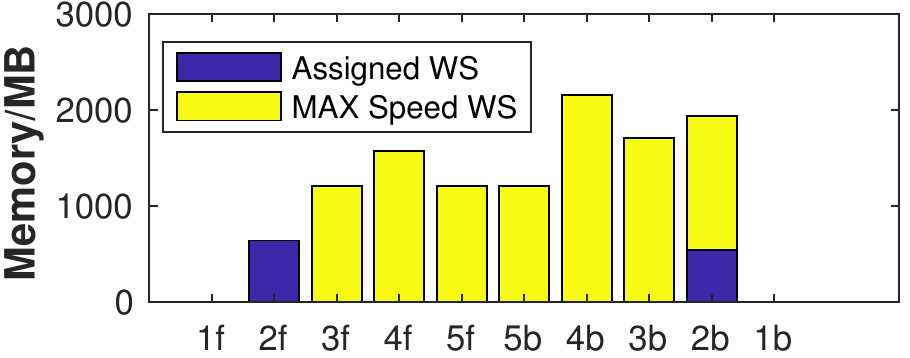}\label{dyna_conv_pool_3}} \quad
\subfloat[][240(imgs/s), pool = 5G]{\includegraphics[width=0.45\columnwidth]{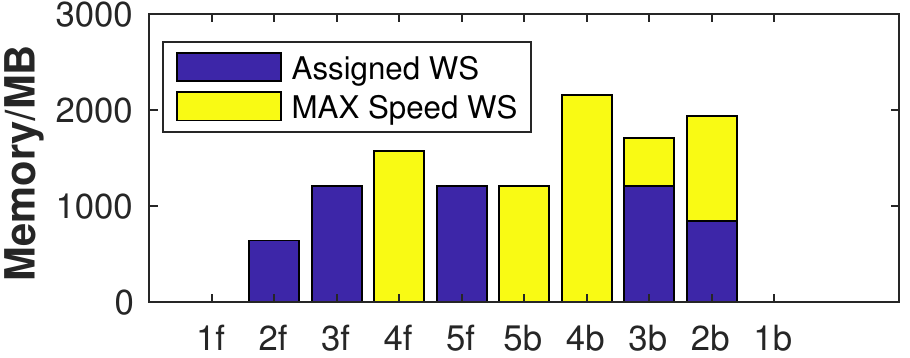}\label{dyna_conv_pool_4}} \quad \\
\caption{Dynamic Conv workspace allocations in the runtime. The digit in x-axis represents the ith CONV layer, while f/d represent the forward and backward. }
\label{dyna_ws}
\end{figure}
 
\textit{\textbf{Dynamic Convolution Workspace Allocation}} intends to optimize each layers' training speed in together with 3 memory techniques. Convolution workspaces are critical to the high performance, while the free memory for convolution workspaces constantly changes at different computing steps as demonstrated in Fig.\ref{recomputation_eval}. The runtime picks the fastest memory-feasible convolution algorithm at a particular step.

Fig.\ref{dyna_conv_batch_2} and Fig.\ref{dyna_conv_batch_1} demonstrate that the runtime 
automatically reduces CONV workspaces to accommodate functional tensors with the increasing batch size. 
Specifically, the runtime prioritizes the functional tensor allocations at batch 300 under 3 GB memory pool (Fig.\ref{dyna_conv_batch_1}), 
while it provisions the most workspace for the maximal speed at batch 100 (Fig.\ref{dyna_conv_batch_2}). 
In general, a higher speed is observable with more convolution workspaces. Fig.\ref{dyna_conv_pool_3} and 
Fig.\ref{dyna_conv_pool_4} demonstrate the training speed (images per second) increases from 203 img/s to 
240 img/s with additional CONV workspaces.

\subsection{Going Deeper and Wider}
Our primary goal is to enable ML practitioners exploring deeper and wider neural architectures within the limited GPU DRAM. 
In this section, we conduct end-to-end comparisions to TensorFlow, MXNet, Caffe and Torch with several mainstream 
linear networks (AlexNet, VGG16) and non-linear ones (ResNet50 $\rightarrow$ 150, Inception V4) under the same experiment setup.

\begin{table}[!t]
\scriptsize
\setlength{\tabcolsep}{0.25em}
\caption{ Going Deeper: the deepest ResNet that different frameworks can reach on a 12GB NVIDIA K40. The batch size is fixed at 16.
ResNet has 4 for-loops to control its depth: $depth = 3*(n_1+n_2+n_3+n_4)+2$, where $n_i$ is the upper limit
of $ith$ for-loop. We fix $n_1 = 6$, $n_2 = 32$, and $n_4 = 6$, while varying $n_3$ to increase the depth. }
\label{deepest_residual}
%\resizebox{200.0pt}{!}{%
\begin{tabular}{c c c c c c c c}
  \hline
    \textbf{Depth}           &  Caffe   &  MXNet &  Torch       & TensorFlow & SuperNeurons    & \\   \hline
    ResNet                   &   148    &  480   &  152         &  592       &  1920           &  \\
   \hline
\end{tabular}
\vspace{-0.2cm}
\end{table} 

\begin{table}[!t]
\scriptsize
\setlength{\tabcolsep}{0.25em}
\caption{ Going Wider: the largest batch size that several mainstream neural architectures
can reach in different frameworks with a 12GB NVIDIA K40. }
\label{largest_batch_all_networks}
\begin{tabular}{c c c c c c c c}
  \hline
    \textbf{peak batch}            &  Caffe   &  MXNet &  Torch       & TensorFlow & SuperNeurons    & \\   \hline
    AlexNet                        &   768    &  768   &  1024        &  1408      &  1792           &  \\
    VGG16                          &   48     &  64    &  48          &  80        &  224            &  \\
    InceptionV4                    &   16     &  N/A   &  N/A         &  64        &  240            &  \\
    ResNet50                       &   24     &  80    &  32          &  128       &  384            &  \\
    ResNet101                      &   16     &  48    &  16          &  80        &  256            &  \\
    ResNet152                      &   16     &  32    &  16          &  48        &  176            &  \\ 
   \hline
\end{tabular}
\end{table}

\begin{figure}[!t]
\centering
\includegraphics[width=0.95\columnwidth]{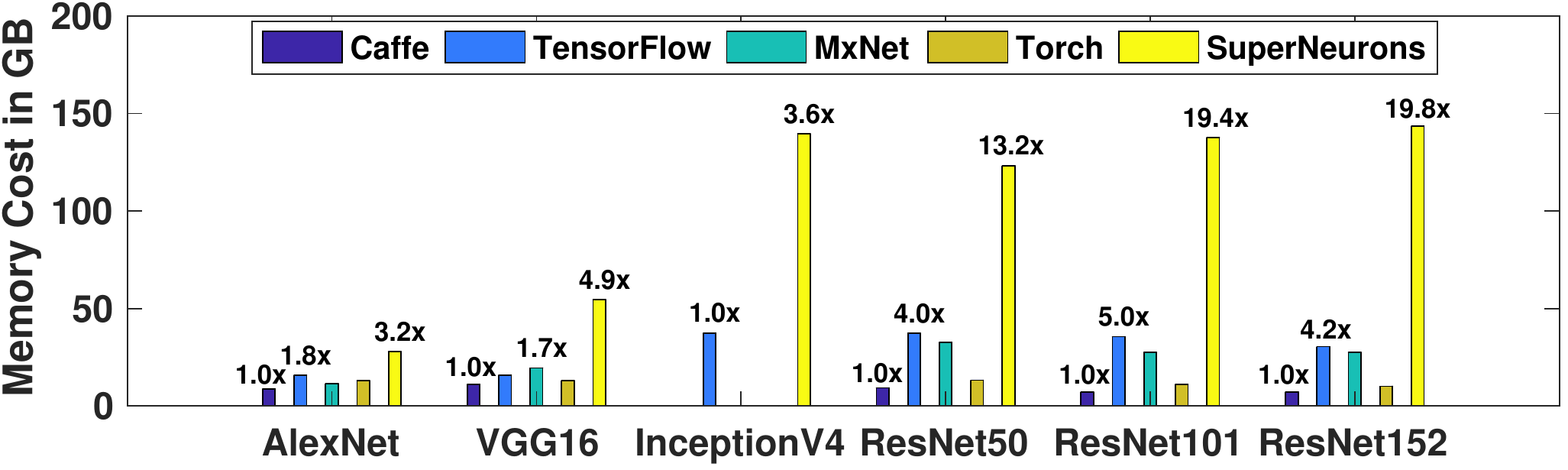}
\caption{ Going Wider: the corresponding memory usages for the batch size in TABLE.\ref{largest_batch_all_networks} }
\label{memory_costs_batches}
\vspace{-0.5cm}
\end{figure}
We increase the batch size to go wider. Table.~\ref{largest_batch_all_networks} presents
the largest batch reachable by different frameworks before the GPU out-of-memory error.
SuperNeurons consistently outperforms the mainstream frameworks on both linear and non-linear networks. 
On average, it handles 1.8947x larger batches than the second best. SuperNeurons can train ResNet101 at 
the batch of 256, which is 3x larger than the second best TensorFlow.

Fig.\ref{memory_costs_batches} demonstrates the corresponding memory requirement to peak batches in Table.\ref{largest_batch_all_networks}. 
The translation is non-linear because of the convolution workspace. We calculate the memory requirement with 
$\sum_{i=1}^{N} l_i^f + \sum_{i=1}^{N} l_i^b$, and $l_i$ is the sum of the memory usages of all tensors in the layer.   
It is observable that SuperNeurons handles up to 19.8x larger model than Caffe. 

% \textcolor{red}{TODO do not use memory usage in figure 13 y axis and caption,
% use memory requirement can be handled. memory usage sounds like we use a lot of memory}

We add layers to go deeper. Table.\ref{deepest_residual} demonstrates SuperNeurons trains 12.9730x, 12.6316x, 4.0000x,
and 3.2432x deeper ResNet than Caffe, Torch, MXNet, and TensorFlow, respectively. Particularly SuperNeurons 
can train a ResNet up to 2500 residual units having approximately $10^4$ basic layers at the batch size of 1
on a 12GB GPU.

%AlexNet Caffe 768, TensorFlow 1408, MXNet 1024,  Torch 1024, SuperNeurons 1792
%caffe tf MXNet torch 64, 80, 112, 80, 224
%32, 128, 112, 32, 384
%16, 80, 64, 16, 256
%80, 32, 240
 
\begin{figure}[t]
\subfloat[][AlexNet]{\includegraphics[width=0.45\columnwidth]{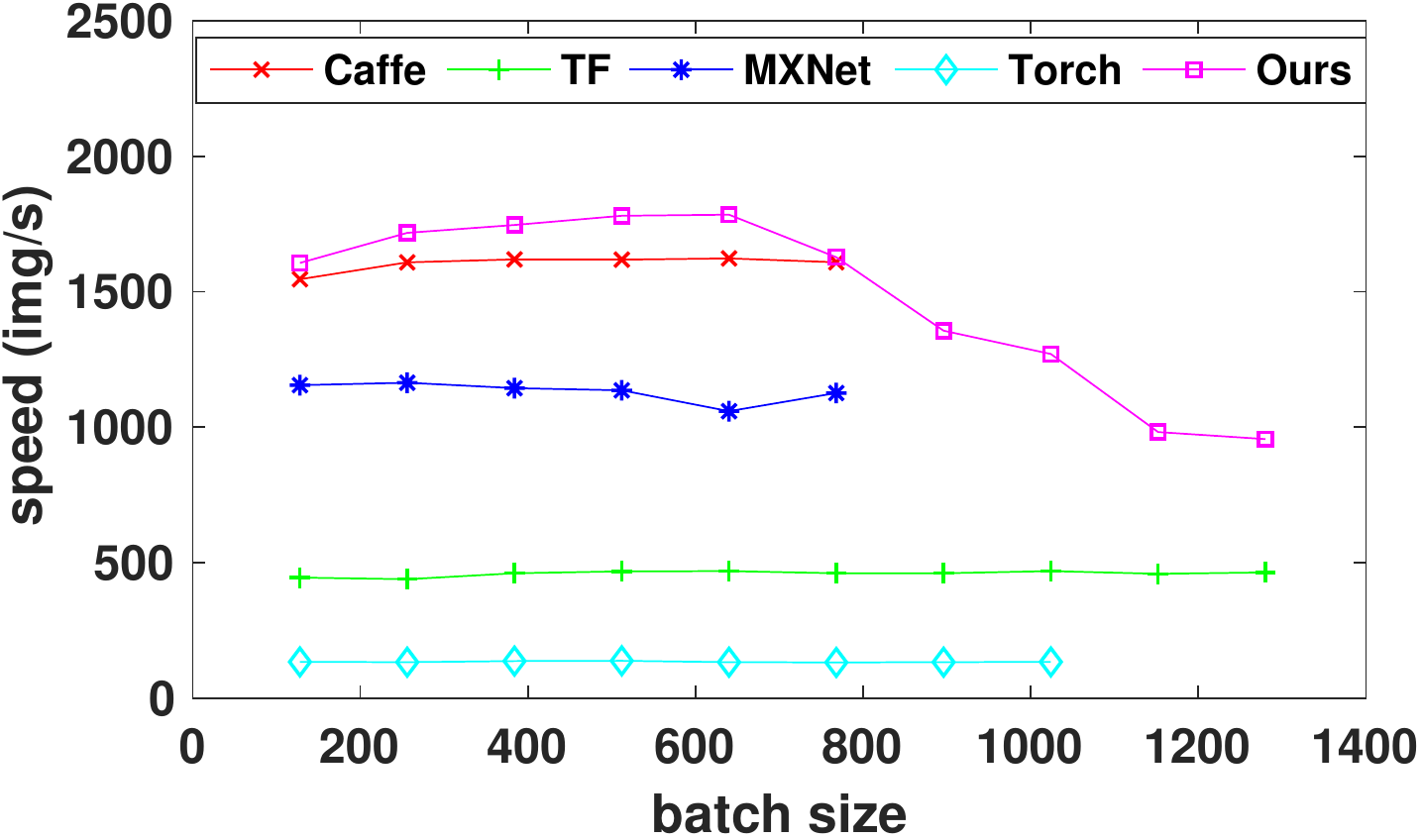}\label{alexnet_speed}} \quad
\subfloat[][ResNet50]{\includegraphics[width=0.45\columnwidth]{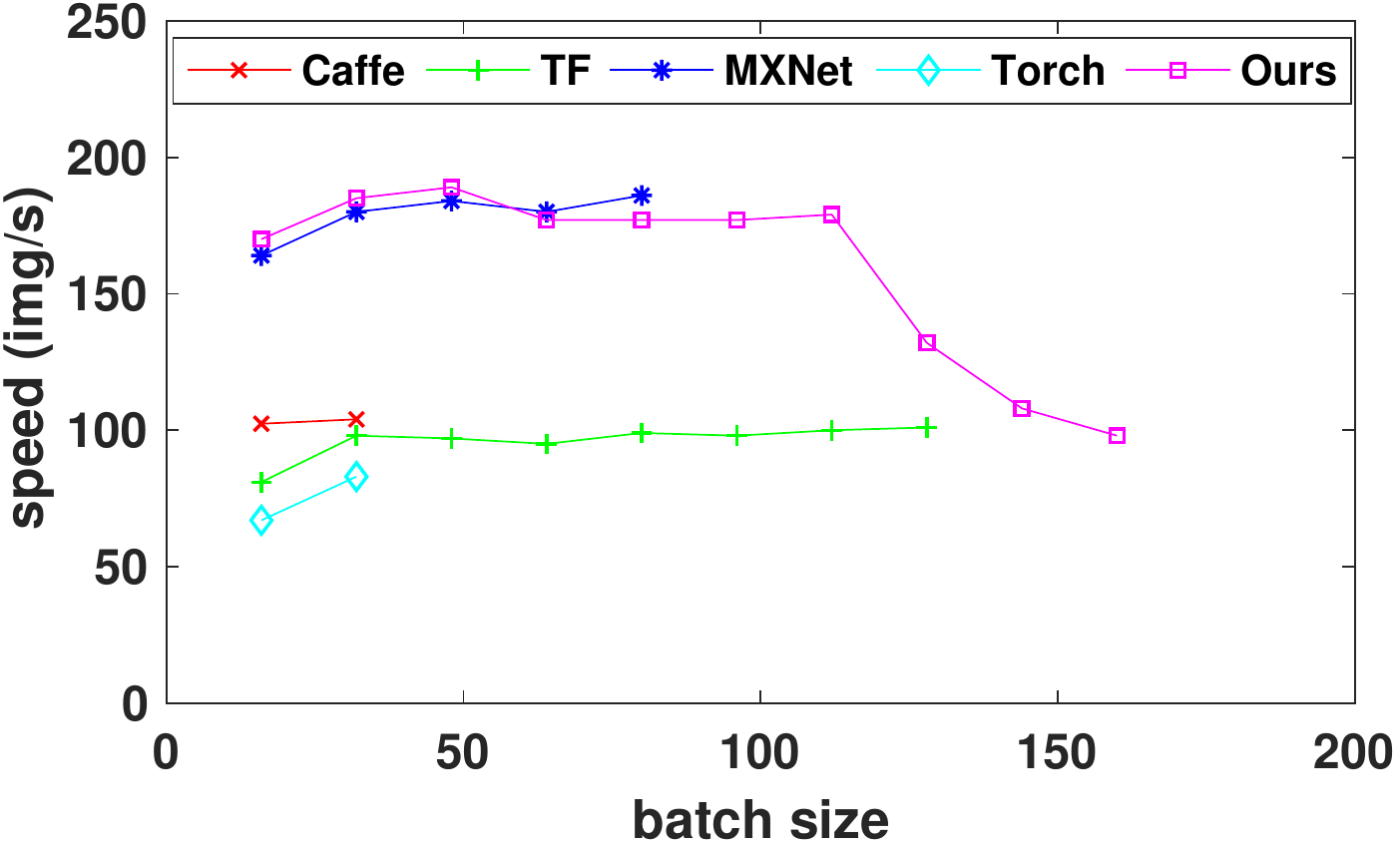}\label{mem_per}} \quad   \\
\subfloat[][VGG16]{\includegraphics[width=0.45\columnwidth]{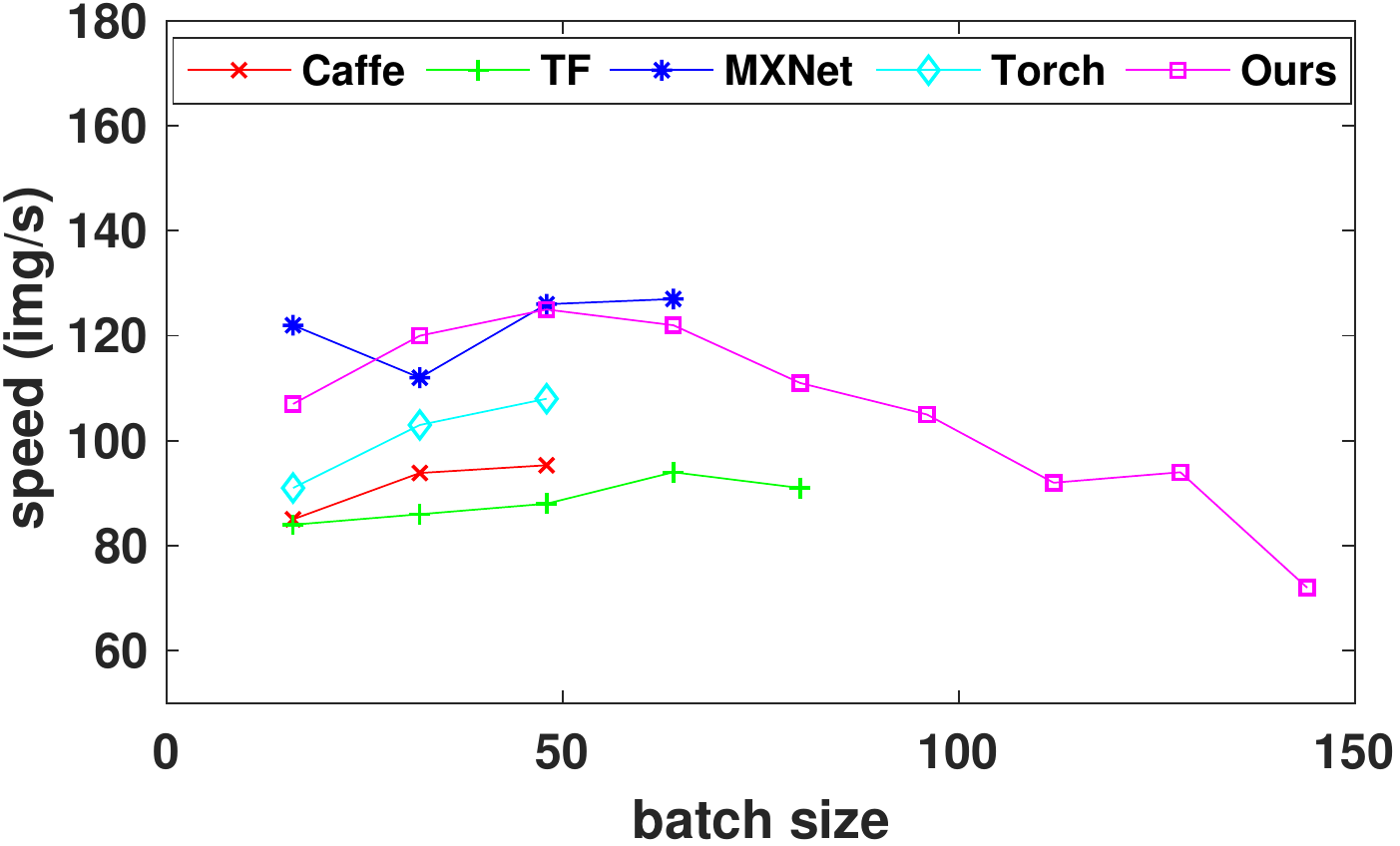}\label{mem_per}} \quad
\subfloat[][ResNet101]{\includegraphics[width=0.45\columnwidth]{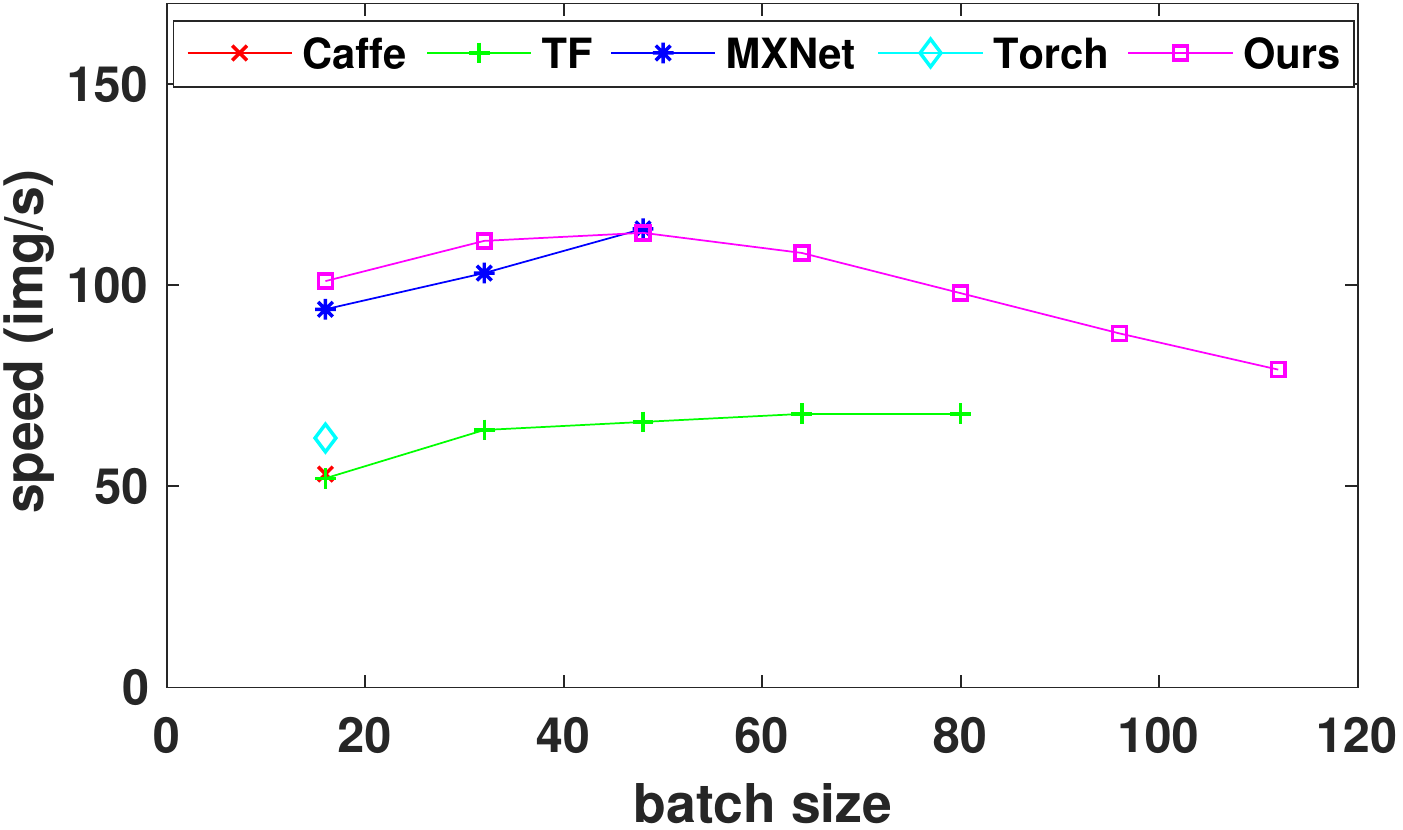}\label{mem_per}} \quad \\
\subfloat[][InceptionV4]{\includegraphics[width=0.45\columnwidth]{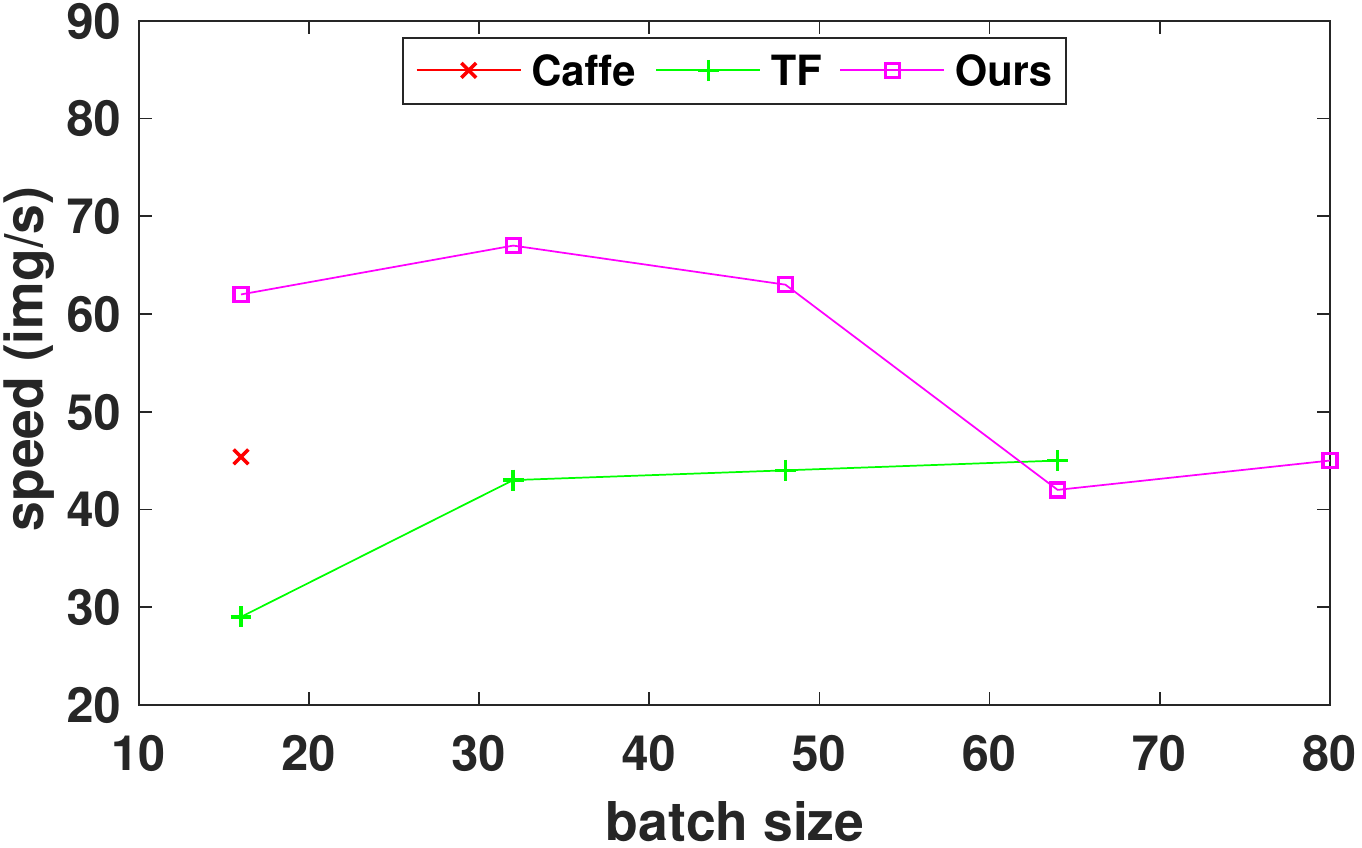}\label{mem_per}} \quad
\subfloat[][ResNet152]{\includegraphics[width=0.45\columnwidth]{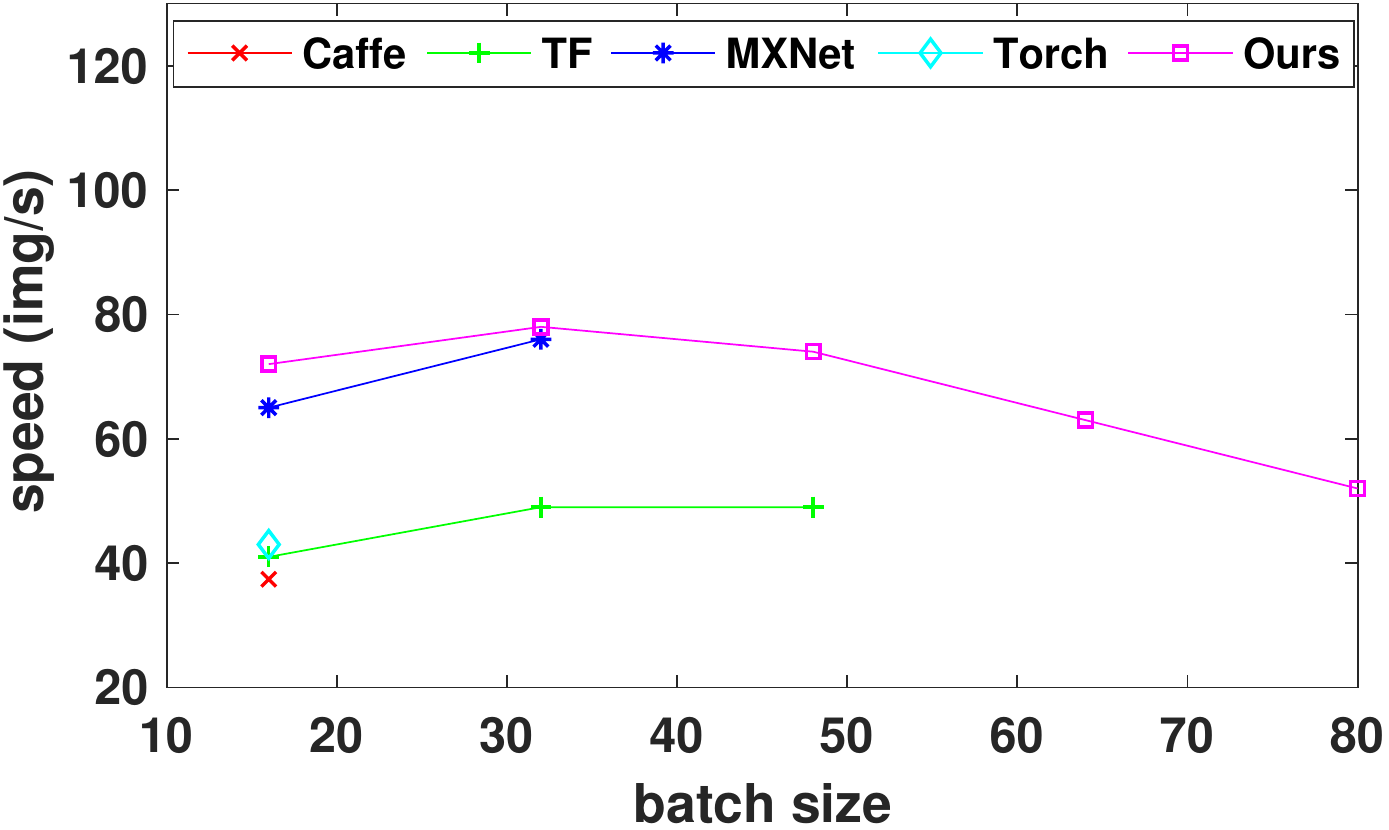}\label{mem_per}} \quad \\
\caption{An end-to-end evaluation of different DL frameworks. We benchmark the data on a TITAN XP.}
\label{overall_speed_eval}
\vspace{-0.5cm}
\end{figure}

%Section \ref{speed_opt} presents several techniques to improve the training speed of our runtime.
%Most importantly, it optimizes the CONV workspace allocations to seek the optimal speed.

The training speed is measured by the processed images per second. Fig.\ref{overall_speed_eval}
presents an end-to-end training speed comparison of SuperNeurons to mainstream DL systems. 
SuperNeurons consistently demonstrates the leading speed on various linear networks (AlexNet, VGG16) and 
nonlinear ones (ResNet$50\rightarrow152$, Inception V4). The performance largely results from the abundant 
supply of convolution workspaces saved by the dynamic GPU memory scheduler. We can also observe that the speed
has slowly deteriorated along the increasing batch size. This is because the growing communications in more 
frequent tensor swapping between CPU and GPU DRAM. The performance will be the worst when GPU memory can only
accommodate one network layer. Then, the runtime has to constantly offload the current layer before
proceeding to the next one.

\section{Related Work}
Several solutions have been proposed to address the GPU DRAM shortage for training large-scale neural networks. Model Parallelism provides a straightforward solution to the large network training. 
DistBelief \cite{dean2012large} partitions a network across multiple machines so that
each machine holds a segment of the original network. 
Coates et al \cite{coates2013deep} discuss another partition scheme on multi-GPUs. 
Model Parallelism demands huge intra-network communications for synchronizations. 
Therefore, most DL systems parallelize the training with Data Parallelism 
for the high-performance \cite{jia2014caffe, abadi2016tensorflow, chen2015MXNet, collobert2002torch}.
In this paper, we focus on the GPU DRAM shortage issue for Data Parallelism.

Under Data Parallelism, vDNN \cite{rhu2016vdnn} proposes a prefetching and offloading technique
to utilize the CPU DRAM as an external buffer for the GPU. 
It tries to overlap communications with computations by asynchronously swapping the data between CPU and GPU 
amid the back-propagation. The performance of this method largely depends on the communication/computation 
ratio. Some layers such as POOL are very cheap to compute, while the GPU processing speed is several orders of 
faster than PCI-E 16x bus. In nonlinear networks, the performance will quickly deteriorate once computations are 
inadequate to overlap with communications. Chen et al \cite{chen2016training} also introduce a recomputation 
strategy to trade computations for memory. However, their method fails to fully exploit the memory saving
opportunities and computation efficiency for ignoring the memory variations among layers.

Removing the parameter redundancy also reduces the memory usage.
For example, the network pruning \cite{han2016eie, hassibi1993second} removes near zero parameters; 
and quantization \cite{vanhoucke2011improving} or precision reduction \cite{judd2016proteus} 
utilize low precision floats to save the memory. Although the parameter reduction
has immense benefits in deploying neural networks on embedded systems, parameters only account for a negligible portion of memory usage in the training. Therefore, these approaches are quite limited to the training.

%
%models the back-propagation with a static computation graph 
%to enable the memory sharing among intermediate results that are no longer needed. However, 
%their discussion is limited to the linear network that each layer is sequentially connected.
%The structure of excelling networks, such as Deep Residual \cite{he2016deep}, DenseNet \cite{huang2016densely},
%or inception v4 \cite{szegedy2017inception} have jumping or fan-out connections rendering 
%non-linear computation patterns. We proposed a new liveness analysis approach with non-linear
%network architecture considered. Chen's model also assumes the homogeneous memory usage on
%network layers, while our model considers the intrinsic difference among network layers by
%leveraging the computation intensity and the communication volume in a layer. vDNN also 
%does not discuss caching data on GPU DRAM to avoid unnecessary data transfers. 

\section{Conclusion}
In this paper, we focus on the GPU memory scheduling problem for training deep neural networks; 
and we propose a novel dynamic scheduling runtime to tackle the issue. The
runtime features three memory techniques to reduce $peak_m$ to $max(l_i)$, 
which is the minimal at the layer-wise granularity. We also
propose several performance optimizations to guarantee the high performance. 
Evaluations against state-of-the-art DL frameworks have demonstrated the effectiveness and efficiency 
of proposed dynamic scheduling runtime. It creates new opportunities for DL practitioners to explore 
deeper and wider neural architectures; and the new accuracy record is awaiting to be refreshed 
with even deeper and wider designs.

\section{Acknowledgements}
This research is funded in part by the DARPA Award 16-43-D3M-FP-040 and gifts from Google, VMware, Mellanox, and Oracle.
Zenglin Xu and Jianmian Ye were supported by a grant from the Natural Science Foundation of China (No. 61572111), 
and a Fundamental Research Fund for the Central Universities of China (No. ZYGX2016Z003).

\bibliographystyle{acm} 

\bibliography{ref}

\end{document}